\documentclass[journal=jacsat,manuscript=article]{achemso}

\usepackage[version=3]{mhchem} 
\usepackage{siunitx}

\author{Michael Verhage}
\affiliation{Eindhoven University of Technology - M2N, Netherlands}
\author{Tunç H. Çiftçi}
\affiliation{Eindhoven University of Technology - FNA, Netherlands}
\author{Michiel Reul}
\affiliation{Eindhoven University of Technology - M2N, Netherlands}
\author{Tamar Cromwijk}
\affiliation{Eindhoven University of Technology - FNA, Netherlands}
\author{Thijs J.N. van Stralen}
\affiliation{Eindhoven University of Technology - FNA, Netherlands}
\author{Bert Koopmans}
\affiliation{Eindhoven University of Technology - FNA, Netherlands}
\author{Oleg Kurnosikov}
\affiliation{University of Lorraine, Institut Jean Lamour, France}
\author{Kees Flipse}
\affiliation{Eindhoven University of Technology - M2N, Netherlands}
\email{c.f.flipse@tue.nl}

\title
  {Switchable-magnetisation planar probe MFM sensor}

\begin{document}


\begin{tocentry}
        \includegraphics{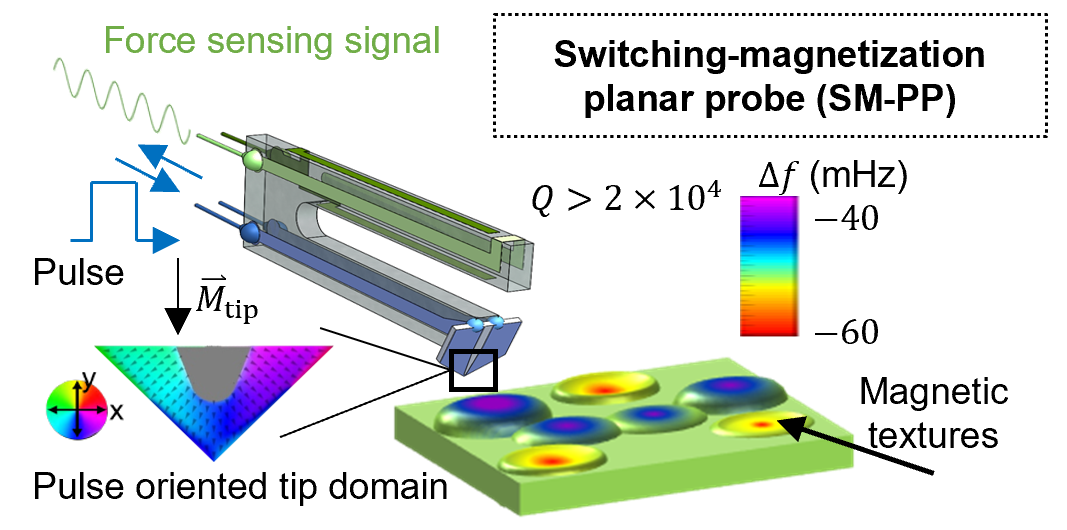}
            \texttt{\textbf{For Table of Contents Only}}
                \label{fig:Tip_Concept}
\end{tocentry}




\begin{abstract}
  We present an alternative switching-magnetization magnetic force microscopy (SM-MFM) method using planar tip-on-chip probes. Unlike traditional needle-like tips, the planar probe approach integrates a microdevice near the tip apex with dedicated functionality. Its \SI{1}{\milli\meter} $\times$ \SI{1}{\milli\meter} planar surface paves the way for freedom in ultra thin-film engineering and micro-/nano-tailoring for application-oriented tip functionalization. Here, we form a microscale current pathway near the tip end
  to control tip magnetisation. The chip like probe or planar probe, was applied to study the complex magnetic behaviour of epitaxial transition metal oxide perovskite LaMnO$_3$, which was previously shown to behave as complex material with domains associated with superpara-, antiferro- and ferromagnetism. To this end we successfully imaged an inhomogeneous distribution of weak ferromagnetic islands with a resolution better than \SI{10}{\nano\meter}.
\end{abstract}


\subsection{Introduction}

Magnetic force microscopy (MFM) is a widespread method in fundamental surface studies and nanoscale technological applications with a high lateral resolution of up to tens of nanometers and \SI{}{\pico\newton} force sensitivity\cite{Kazakova2019FrontiersMicroscopy, Hug1998QuantitativeSamples}. The working principle of MFM relies on the force interaction between the tip's magnetic stray field and a samples' spatially varying magnetic textures. By nanoscale utilization of this magnetic force interaction, MFM covers a wide operational range from characterization to manipulation of magnetic objects \cite{Casiraghi2019IndividualGradients, Albisetti2016NanopatterningLithography, Gartside2018RealizationWriting}.  

Despite its extensive use, a general MFM hits its capability limits mainly in lateral resolution of imaging materials with weak or time-varying magnetization. For instance, low coercive and weak ferromagnetic (FM) or superparamagnetic (SP) structures generally require an external magnetic field to saturate \cite{Schreiber2008MagneticNanoparticles}, and without this the magnetic force to the tip is weak or may be undetectable by the bandwidth of the MFM. Hence, nonmagnetic interactions such as those of electrostatic origin can mask the magnetic signal\cite{Torre2011MagneticNanoparticles, Angeloni2016RemovalNanoparticles, Krivcov2018UnderstandingResolution}. To obtain the pure magnetic signal of nanoscale weak FM or SP textures, such as isolated islands, an MFM variant called switching magnetization force microscopy (SM-FM) \cite{Cambel2011SwitchingMFM, Cambel2013HighMicroscopy} or controlled magnetization-MFM (CM-MFM) \cite{Angeloni2016RemovalNanoparticles} stands out by extracting such signals out of the detected force \cite{Cambel2013HighMicroscopy, Wren2017SwitchableStructures, Krivcov2018UnderstandingResolution}. Beyond traditional MFM, SM-FM measures a relative force change due to controlled altering of the magnetic state of the tip or the sample (or both). Only the magnetic field interaction is sensitive to the relative magnetic polarities between the tip and sample and hence can thus be detected. 

The need for a SM-FM imaging technique with a capability of imaging weak FM or SP islands with a resolution beyond \SI{10}{\nano\meter} can be found in the study of epitaxial complex oxide perovskites such as LaMnO$_3$ (LMO$_3$). Wang \textit{et al.} \cite{Wang2015ImagingHeterostructures} have shown that epitaxial LMO$_3$ reveals an abrupt transition from an antiferromagnetic (AF) state to a ferromagnetic (FM) depending on the thickness of the LMO$_3$ layer. The magnetic transition occurred at a film of \SI{5}{} atomic unit cell (u.c.) thickness. Furthermore, Anahory \textit{et al.}\cite{Anahory2016EmergentInterfaces} observed inhomogeneously distributed SP islands besides the FM domains, with the former only detected following an applied in-plane magnetic field of variable strength. Both groups used a scanning SQUID microscope (SSM), albeit with different lateral resolution, to image the LMO$_3$ sample's stray field distribution. However, the SSM imaging performed by Anahory \textit{et al.} \cite{Anahory2016EmergentInterfaces} could not go beyond a resolution of \SI{100}{\nano\meter}, which left the SP island size to be only indirectly inferred between \SI{10}{\nano\meter} and \SI{20}{\nano\meter}. 

To solve the problem of limited imaging resolution of traditional MFM, we carefully designed a new type of SM-FM sensor. To this end, we design a magnetic tip with a stray-field of several hundred mT strong enough to saturate the magnetic textures. The tip is realised by forming an oriented single domain state near the tip apex \cite{Corte-Leon2019MagneticWalls}. Traditional needle-like MFM tips generally only generate up to a few tens of mT of stray field \cite{Sakar2021QuantumMicroscopy, Hug1998QuantitativeSamples}. With this approach the LMO$_3$ weak FM domains are simultaneously saturated and profiled for imaging by the same tip. The tip's stray field decays rapidly from the tip and hence, by changing the height of the tip with respect to the sample surface the weak magnetic textures can be actively saturated. 

\begin{figure*}
    \centering
        \includegraphics[scale=0.4]{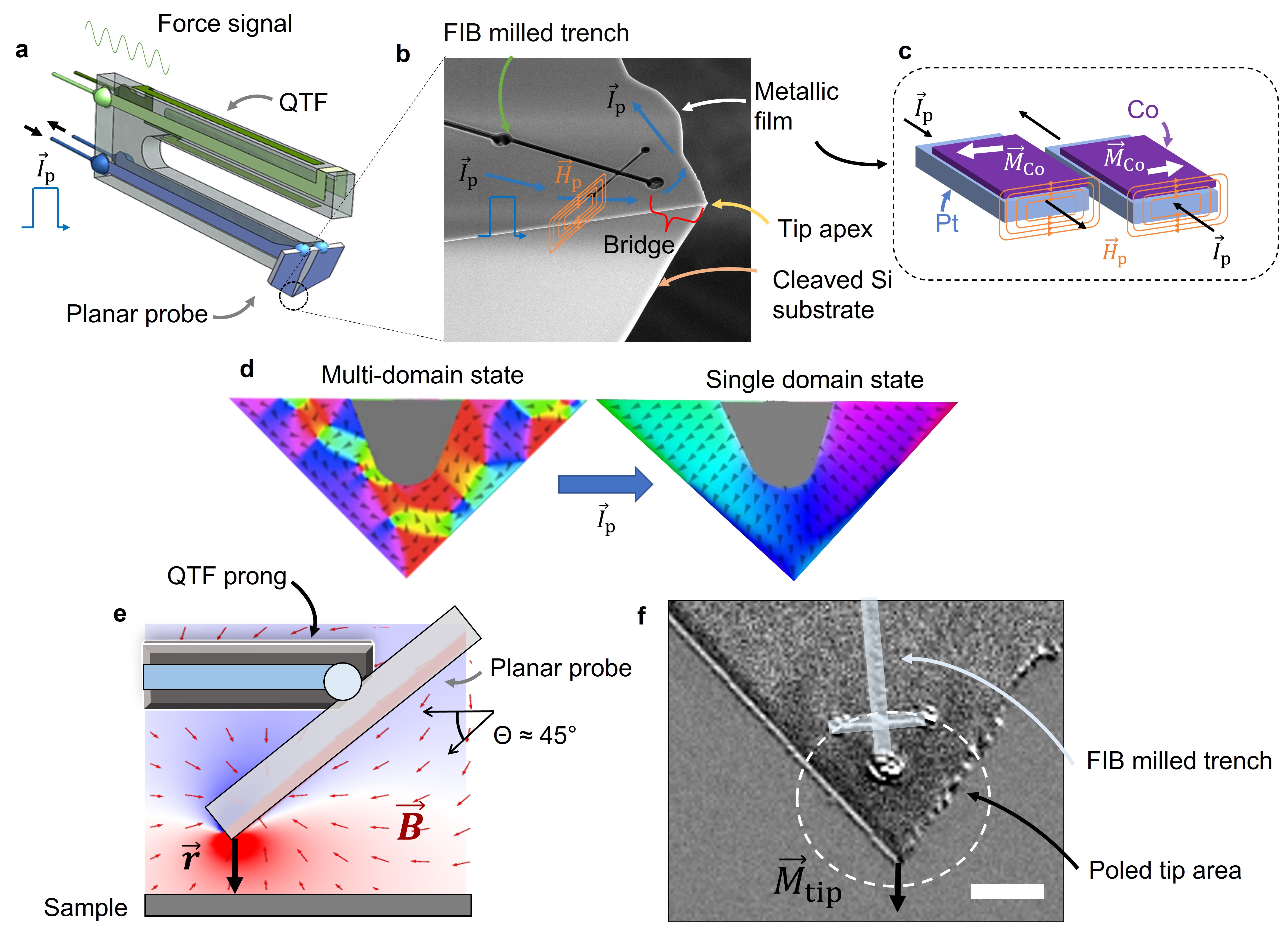}
            \caption{\textbf{Planar probe with switchable tip magnetization (SM-PP).} (\textbf{a}) Illustration of the SM-PP sensor, with electrical contacts for force sensing  and sending current pulses $I_{p}$ to the tip apex. The tip is formed by a tip-on-chip called the planar probe (PP). (\textbf{b}) SEM image of the PP with a sharp tip apex formed by cleaving a Si wafer. Sending a current pulse generates an \O rsted field ($\Vec{H_{p}}$) within the metallic film to orient the tip magnetization into a singular domain state. The current pathway is created by the formation of a FIB milled bridge. (\textbf{c}) The metallic film with two main layers: the current-carrying Pt layer and the ferromagnetic Co layer. The polarity of $I_{p}$ determines the direction of the \O rsted field ($\Vec{H_{p}}$) which alters (reverse) the direction of magnetization of the Co film. (\textbf{d}) Multi-domain state can be poled into an oriented single domain by a controlled current pulse. (\textbf{e}) Schematic side view of the SM-PP shows the mass retuned tuning fork prong and a lateral view of the surrounding tip stray field $\vec{B}$. The planar probe is placed under a \SI{45}{\degree} angle to the prong. (\textbf{f}) Kerr microscopy image showing a poled magnetic tip domain after sending a single $I_{p}$. The dark contrast at the bottom of the tip demonstrates the singular domain with magnetization $\Vec{M_{\text{tip}}}$. The white scale bar equals \SI{5}{\micro\meter}.}
            \label{fig:TipConcept}
\end{figure*}

We demonstrate that our sensor is capable of imaging magnetic textures of LMO$_3$ with a resolution beyond \SI{10}{\nano\meter}. For this, we present a new approach combining planar chip-like probes\cite{Siahaan2015CleavedMicroscopy, Ciftci2019PolymerProbes, LeeuwenhoekFabricationMicroscopy} with highly sensitive tuning fork force sensors which we call the switching-magnetization planar probe (SM-PP) and is illustrated in Figure \ref{fig:TipConcept}a. This method aims to provide an on-chip reorientable tip magnetization with no required external magnetic field, to act as an switchable magnetic force sensor. 


\subsection{Results and Discussion}

The working principle of the SM-PP relies on switching from a multi-domain state of the tip to a poled single domain via an internally generated \O rsted field ($H_{p}$) within a planar chip-like probe, as illustrated in Figures \ref{fig:TipConcept}a, b and d. Initially, the magnetic layer on the tip is in a multi-domain state with a closed flux loop, Figure \ref{fig:TipConcept}d. The direction of this flux may be irregular, and hence inappropriate for perturbing weak FM islands. The planar probe design used in this study has a bi-metallic structure of thin-film components: the current-carrying layer and the ferromagnetic layer, as depicted in Figures \ref{fig:TipConcept}b and c. By sending an electrical pulse ($I_{p}$) through the current-carrying layer in a designated electrical pathway (called the bridge) near the tip apex, we generate an \O rsted field of controlled magnitude and well-defined direction which penetrates the ferromagnetic layer. This action leads to a singular domain state of the tip apex, with a preferable orientation. 

The planar probe is formed by cleaving a silicon wafer into a small \SI{1}{\square\milli\meter} square piece with a \SI{90}{\degree} tip apex \cite{Ciftci2022EnhancingProbes, Siahaan2015CleavedMicroscopy}. Near the tip apex, i.e. the cleaved corner, naturally increases the flux density, increasing the tip stray field compared to a needle-like MFM tip. This magnetic field strength and distribution is discussed later on. The single domain state of the tip can be used to probe weak FM domains. To this end, we used a \SI{30}{\nano\meter} Pt film for the current-carrying layer and a \SI{15}{\nano\meter} Co film for the ferromagnetic layer, placed on top of the planar probe. Detailed fabrication procedures of the film and planar probe are given in Supplementary S1. 

Contrary to the traditional passive needle-like MFM tips, we can orient the SM-PP multi-domains into a singular domain by only a single current pulse as often as needed to combat transient tip demagnetisation, a known issue in MFM. The resulting tip domain after sending a current pulse is illustrated in Figure \ref{fig:TipConcept}d. As a result, we can obtain consistently oriented tip domains, resulting in a predictable stray field in the tip vicinity, as indicated by $\vec{B}$ in Figure \ref{fig:TipConcept}e. Figure \ref{fig:TipConcept}e illustrates the side view of the SM-PP with the tip stray field distribution predominately out-of-plane from the sample's perspective. In Supplementary S4 we discuss in-depth the tip stray field distribution derived from a numerical study. Finally, Figure \ref{fig:TipConcept}f
shows a Kerr microscopy image of the SM-PP, after having sent a current pulse of sufficient amplitude. A poled tip domain is formed as observed with the dark contrast near the apex, highlighted within the dashed circle. 

We attached this functionalized planar probe to a mass retuned \cite{Ciftci2022EnhancingProbes} quartz tuning fork (QTF) force sensor with integrated electrical access to the probe for the current pulse $I_{p}$, as schematically illustrated in Figure \ref{fig:TipConcept}a.  QTF's have been successfully used before for MFM \cite{Schneiderbauer2012QPlusResolution} and are easily integrated in an UHV scanning probe microscope. The retuned tuning fork approach significantly improved the load capacity of the QTF sensor. As widespread AFM and MFM applications have previously experienced, once the mass exceeds several tens of \SI{}{\micro\gram} \cite{Dagdeviren2017OptimizingAnalysis}, which is far below the mass of a chip-like probe, the oscillation's Q-factor value drops to only several hundred from the original 40 thousand. This reduction in Q-factor results in a large loss in force sensing capabilities \cite{Ooe2014ResonanceMicroscopy, Ciftci2022EnhancingProbes}. For degraded Q-factor sensors, we would be unable to use the planar probe for imaging magnetic fields of LMO$_3$. To this end, the retuned tuning fork approach compensates for the mass unbalancing from planar probe attachment and recovers sensitivity. As a result we are able to restore the Q-factor to over \SI{2e4}{} at room temperature in ultra-high vacuum (UHV) \cite{Ciftci2019PolymerProbes, Ciftci2022EnhancingProbes}.
In Supplementary S3 we discuss further the need for a high $Q$.

As a consequence, the Q-factor drops to only a few hundred \cite{Ciftci2019PolymerProbes}, leading to degraded force sensing capabilities. The same effect can arise from the additional wires connecting the current control signal for pulsing, which is the reason why dedicated electrical contacts to the tip are now integrated on the tuning fork itself \cite{Giessibl2019TheMicroscope}. We solve the mass-imbalance by mass retuning \cite{Ooe2014ResonanceMicroscopy} the QTF as described in our previous work \cite{Ciftci2022EnhancingProbes} and utilising readily available electrodes on the tuning fork.  

For extracting the magnetic signal of weak FM islands, the SM-PP needs to change to a fully oriented $M_{\text{tip}}$ near the tip apex, starting from a multi-domain state. We turned to finite element modeling (FEM), with COMSOL\texttrademark, to simulate the generated \O rsted field within the bridge  to assess the required $I_{p}$ magnitude for tip magnetisation control. Furthermore, we can investigate the thermal response of the tip by Joule heating.  


\begin{figure*}
    \centering
        \includegraphics[scale=0.45]{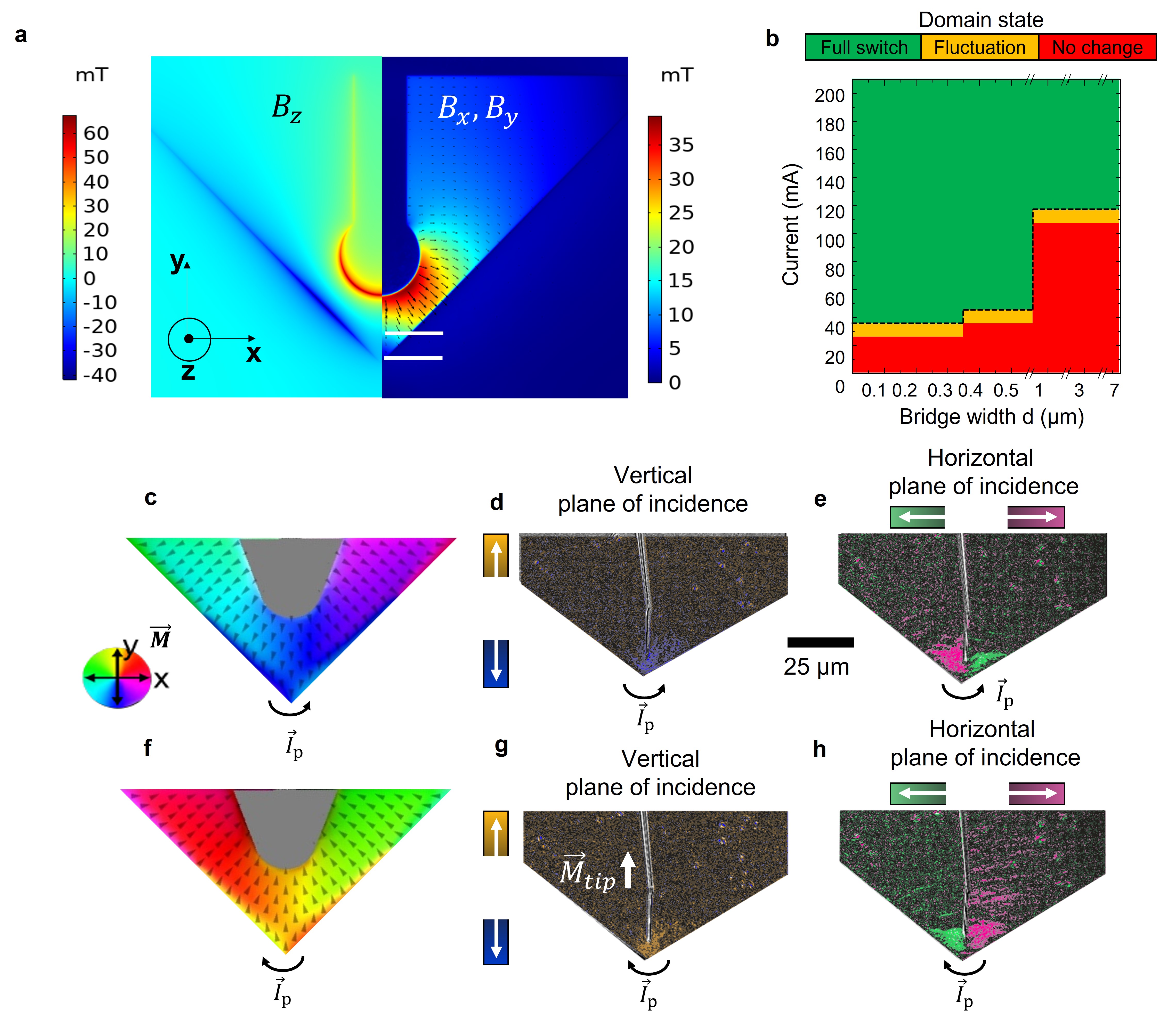}
            \caption{\textbf{Numerical and experimental validation of the magnetic switch of the SM-PP tip.} (\textbf{a}) Numerical calculations of $\vec{H_{\text{p}}}(\Vec{r})$ magnetic field components ($B_x$,$B_y$) and $B_z$) from a \SI{150}{\milli\ampere} pulse.  The bridge is \SI{5}{\micro\meter} wide. (\textbf{b}) Numerical switching behaviour of the SM-PP, covering bridge width $d$ varying from \SI{50}{\nano\meter} to \SI{7}{\micro\meter}, presented in the phase diagram. The colors indicate a switch between two oppositely poled single domain state (green), domain fluctuations (orange) or no switch (red). (\textbf{c}), (\textbf{f}) Simulated single domain formation of the tip for inverting $I_{p}$ polarity. (\textbf{d})-(\textbf{h}) Kerr microscopy results show domain orientation switching after inverting $I_{p}$ polarity. The vertical component of the altered magnetization (indicated in blue and yellow) is visible in the location near the tip end in \textbf{d} and \textbf{g}. The horizontal component is mostly aside the tip end \textbf{e, h}.}
            \label{fig:TipSwitch}
\end{figure*}

Figure \ref{fig:TipSwitch} presents the numerical and experimental validation of the magnetic switch of SM-PP tip. The simulations cover various bridge widths $d$ in the range from \SI{50}{\nano\meter} to \SI{7}{\micro\meter} and different $I_{p}$ values from \SI{10}{\milli\ampere} to \SI{200}{\milli\ampere}. The pulse duration is \SI{500}{\nano\second}. See Supplementary S4 for details on the simulations. First, Figure \ref{fig:TipSwitch}a shows the calculated spatial field components of $\vec{H_{\text{p}}}(\Vec{r})$ of a \SI{5}{\micro\meter} bridge under application of $I_{p}=$ \SI{150}{\milli\ampere}. The in-plane field components $B_x$ and $B_y$ of $\vec{H_{\text{p}}}(\Vec{r})$ follow the bridge structure. This indicates that both symmetric sides of the tip have opposing magnetic direction, as is evident from the current flow pathway. Near the tip apex, $B_x$ and $B_y$ are relatively small since the current density is lowest (between the white lines of Figure \ref{fig:TipSwitch}a). A strong out-of-plane component $B_z$ (Figure \ref{fig:TipSwitch}a) is only observed at the boundary of the tip and bridge, but is of little importance with respect to the in-plane magnetisation of the Co film. At just \SI{1}{\micro\meter} away from the tip apex, above the upper white line, the in-plane magnetic field is larger than \SI{10}{\milli\tesla}  which implies the nucleation of oriented in-plane Co domains. In Supplementary S1 the magnetisation response of the Co film is given.  

Following, $\vec{H_{\text{p}}}(\Vec{r})$ is used as an input parameter within Mumax$^3$ to calculate the magnetisation response (switch vs. no switch) of the Co film at the tip apex, as a function of the bridge width $d$ and $I_{\text{p}}$. In Figure \ref{fig:TipSwitch}b the color scale represents three different states of the tip magnetization after applying $I_{p}$. Green means the tip end domain shows a 180\textdegree\hspace{1pt} reversal, so a full switch. Yellow represents an observed modification or a limited rotation by less than 180\textdegree\hspace{1pt} in the tip domain. Red implies that the magnetization remained identical to the pre-pulse orientation. The results show a few tens of mA increase in critical current level for the bridge gap width values from \SI{50}{\nano\meter} until \SI{1}{\micro\meter}, as given in Figure \ref{fig:TipSwitch}a. For the bridge gap widths greater than \SI{1}{\micro\meter}, the critical current shows a larger increase. 

Although the nanometer scale of the bridge can be achieved with various techniques and types of lithography, in our experiments we used focused ion beam (FIB) milling. This resulted in bridges on the micrometer scale and as the simulations results shows, we require a current magnitude in the order of \SI{e2}{\milli\ampere}. Supplementary S1 discusses the FIB fabrication in further detail. When we simulate a current pulse with \SI{130}{\milli\ampere} amplitude and \SI{500}{\nano\second} duration for a bridge of \SI{5}{\micro\meter}, the tip magnetization changes fully accordingly to the pulse polarity, as shown in Figures \ref{fig:TipSwitch}c and f. The notion of tip switch at values below \SI{130}{\milli\ampere} is important, especially for micrometer scale bridges, because it significantly limits the Joule heating, as we discuss later.  

Based on the simulation results, a Kerr microscopy experiment was performed to validate the magnetic switch. The Kerr microscopy experimental details are given in Supplementary S2. Gray/black tones in Figure \ref{fig:TipSwitch}d, e, g, and h represent Co domains preserving initial orientations before the pulse, upon applying a current pulse. False colored areas represent the Co domains' response to $I_{p}$. Figures \ref{fig:TipSwitch}c and f indicate the orientation in the vertical direction expressed by the blue-to-yellow color scale. Figures \ref{fig:TipSwitch}d and g show the domain orientation in the horizontal direction, given by the pink-to-green color scale. The domain is mainly confined to the bridge region, as only here the current density is sufficient for inducing Co domain reversal. Along the length of the FIB bridge, the domains are inverted (pink and green), which follows closely those of the numerical simulations of Figures \ref{fig:TipSwitch}c and f, validating the realisation of the SM-PP. 

\begin{figure*}
\centering
    \includegraphics[scale=0.5]{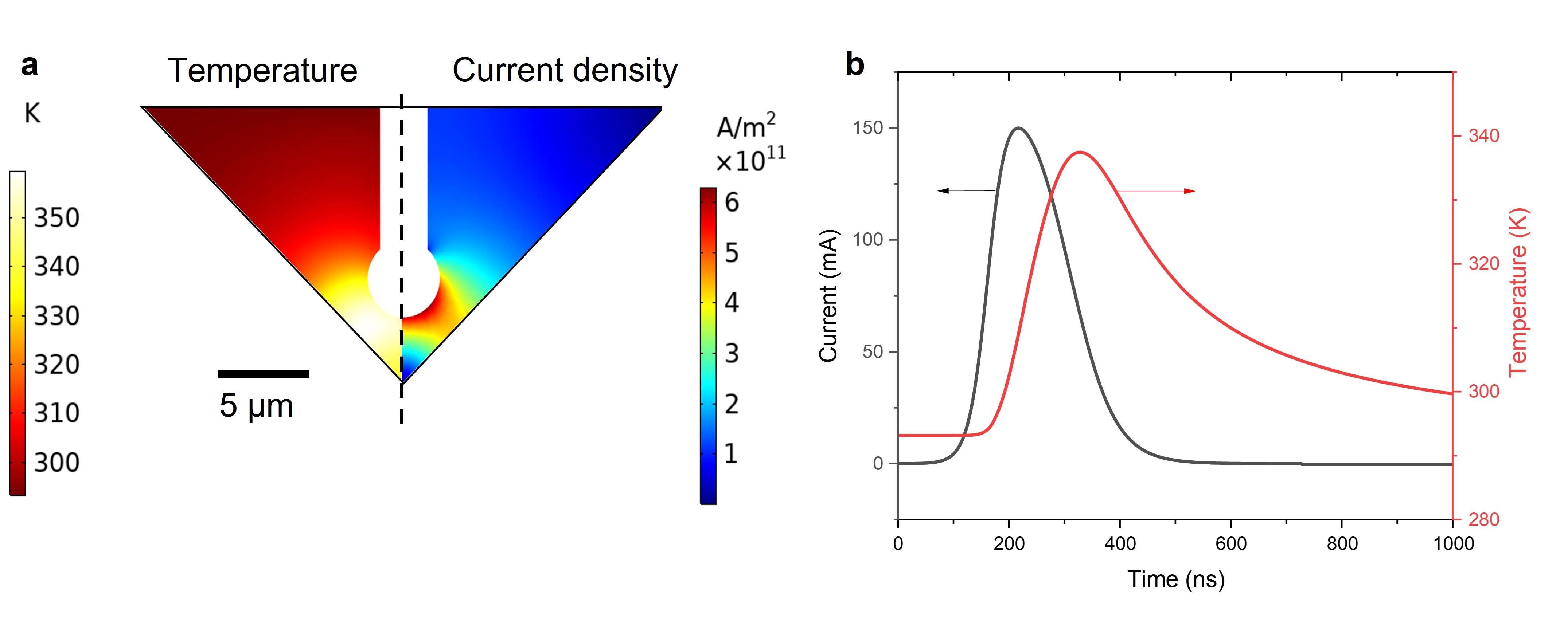}
    \caption{\textbf{Numerical study of the thermal response.} (\textbf{a}) On the right, the calculated current density across the bridge for a current pulse of \SI{150}{\milli\ampere}. The current density increases up to \SI{6e11}{\ampere\per\meter\squared} at the smallest section of the bridge. On the left, the corresponding temperature profile. (\textbf{b}) The current pulse $I_{p}$ has to form of an asymmetric double sigmoidal function plotted as the black curve. With a FWHM of \SI{160}{\nano\second} and a peak value of \SI{150}{\milli\ampere}. The transient temperature response, red curve, show a rapid decay of the temperature, highlighting the efficient thermal dissipation of the bridge and ensuring mechanical stability.}
    \label{fig:Numerical}
\end{figure*}

After applying $I_{p}$, the temperature increase should be excessive i.e. above \SI{100}{\kelvin}, because it would hamper operation in UHV and degrade the tip's metallic layers. Examples and solution by metallic layer composition with respect to preventing degradation of tips are discussed in Supplementary S1. Hence, we modelled the (transient) temperature response of the tip for $I_{p}=$ \SI{150}{\milli\ampere} for an upper limit of thermal increase. Figure \ref{fig:Numerical}a compares the simulated spatial current density across the bridge for a $I_{p}$ of \SI{160}{\nano\second}, with the thermal profile. As expected, the current density is highest near the shortest width of the metallic film and is in the order of \SI{e11}{\ampere\per\square\meter}. Yet, the maximum temperature increase is observed to be only \SI{50}{\kelvin}, which means operation in UHV is possible and would minimize Joule heating damage to the metallic films. We experimentally pulsed several tips for tens of times and no degradation was observed. The transient heating response was also simulated, with the results given in Figure \ref{fig:Numerical}b. Here, a \SI{160}{\nano\second} asymmetric double sigmoidal pulse, see Supplementary S4 for pulse details, is simulated. The temperature decreases quickly within a microsecond due to efficient thermal dissipation of the Si substrate. We studied the effects of substrate capping material, i.e. Si coated with SiO$_2$ or MgO, on this thermal dissipation and the results are also discussed in Supplementary S4.


To conclude the first part of this work; the design, fabrication and optimisation of SM-PP provides us a SM-PP sensor with high Q-factor. Combined with the current-controlled tip magnetization it enables the possibility to study the magnetic surface textures of LMO$_3$ \cite{Wang2015ImagingHeterostructures, Anahory2016EmergentInterfaces}. For the second part of this work we turn to applying the SM-PP sensor to saturate and image the weak FM islands, aswell the AF domains the former are embedded into, of epitaxial LMO$_3$. 

The magnetic texture of a 6 u.c. LMO$_3$ on STO$_3$ sample was imaged with our MFM operating both below ($T=$ \SI{100}{\kelvin}) and above ($T=$ \SI{300}{\kelvin}) of LMO$_3$'s $T_c=$ \SI{115}{\kelvin} \cite{Wang2015ImagingHeterostructures}. The first aim was to identify the AF and weak-FM texture distribution across the surface. Secondly, the SM-PP is able to magnetize magnetic islands by the tip's oriented stray field exceeding \SI{300}{\milli\tesla}, see Supplementary S4, and hence the size of the magnetic islands can be observed with a lateral scale between \SI{10}{} and \SI{20}{\nano\meter} \cite{Anahory2016EmergentInterfaces}. The same SM-PP sensor was used for all imaging, with Frequency Modulation (FM) feedback. The scanning parameters are kept constant throughout all the measurements, see Supplementary S6 for methods and experimental details.

\begin{figure*}
    \centering
        \includegraphics[scale=0.47]{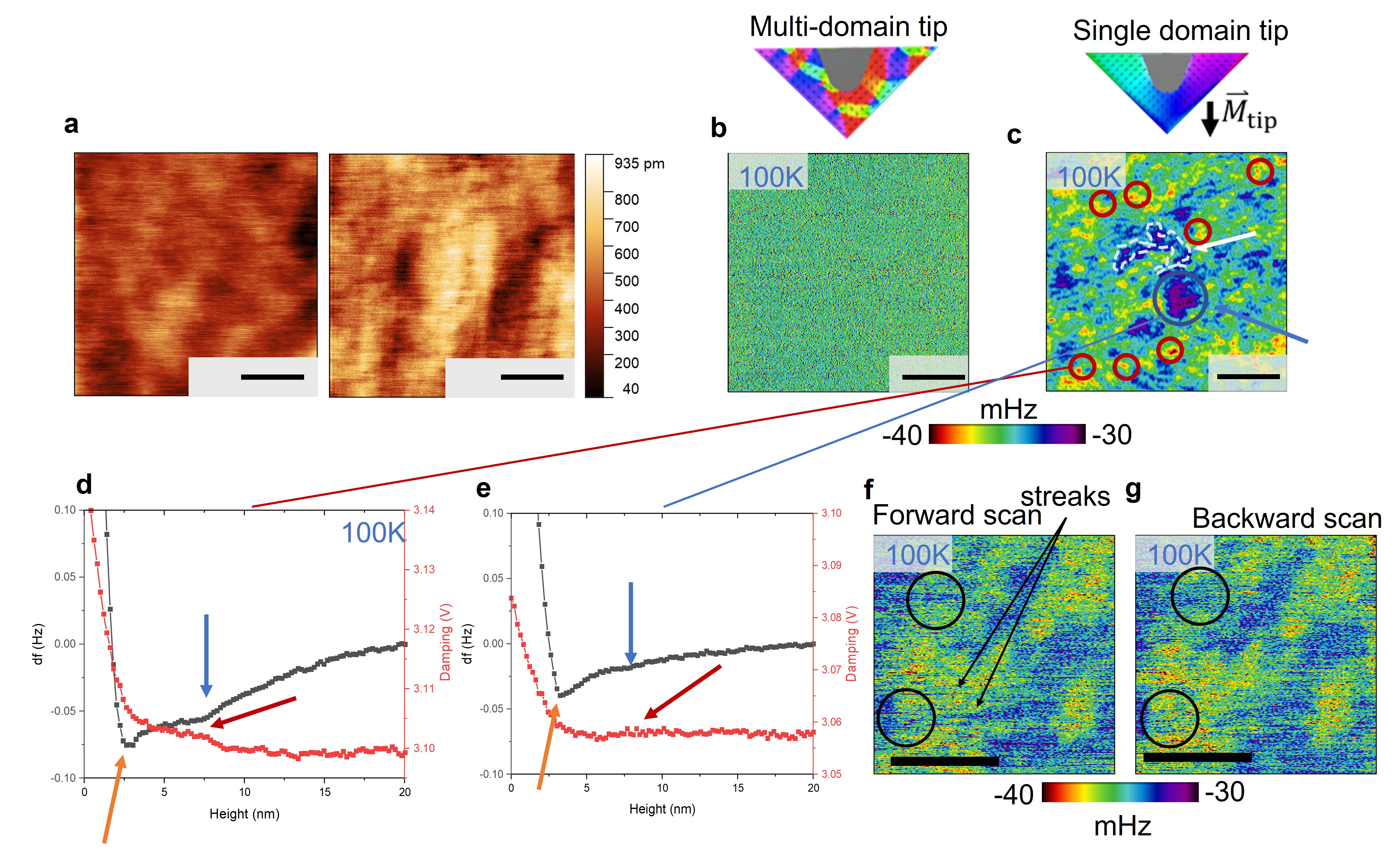}
            \caption{\textbf{MFM images obtained with the SM-PP sensor on 6 u.c. LMO$_3$/STO.} (\textbf{a})  Topographic images of LMO$_3$. (\textbf{b}) Multi-domain tip state MFM measurement at \SI{100}{\kelvin} showing no magnetic contrast. (\textbf{c}) The SM-PP tip is magnetised into a single domain. MFM imaging reveals spatially inhomogenous magnetic contrast at \SI{100}{\kelvin}.  (\textbf{d}) Typical force-distance (F-z) spectroscopy and damping (voltage) signal taken at red areas of \textbf{c}. F-z spectroscopy shows a sudden kink in the attractive regime as highlighted with the blue arrow. The orange arrow indicated short range vdW forces. The damping signal (red line) is simultaneously taken, showing a sudden change in dissipation as indicated with the red arrow, (\textbf{e}) F-z spectroscopy and damping signal taken at a blue spot of \textbf{c}, showing significant reduction in sudden dissipation and force changes at the blue and red arrow, compared to \textbf{d}. (\textbf{f}, \textbf{g}) Magnetic features observed with a poled tip. The tip's stray field induces local magnetic domain perturbation (streaks) as indicated with the black circles and arrows. The forward and backward scan are compared. In all images the black scale bar is equal to \SI{30}{\nano\meter}.}
                \label{fig:MFM}
\end{figure*}

First, we scanned at a temperature of \SI{100}{\kelvin} and imaged a plateau of the LMO$_3$/STO$_3$ stepped surface. The 90$\times$90 \SI{}{\square\nano\meter} topography images are given in Figure \ref{fig:MFM}a and demonstrates a LMO film RMS roughness $S_q$ of \SI{32}{\pico\meter}. Although the surface of LMO$_3$ can show up to 1 u.c. roughness variations, in-homogeneously distributed across the stepped surface, which is a known surface feature for manganites \cite{Gambardella2014SurfaceFilms}. The lateral resolution in topography is limited by the relatively large amplitude of \SI{10}{\nano\meter} used for detecting the long range magnetic force. Ideally, one would use a small amplitude for high resolution topography and a large amplitude for lift mode magnetic imaging \cite{Schneiderbauer2012QPlusResolution}. We leave this for future work, as currently this approach of consecutively switching of the amplitude introduced large drift in our setup. 

After obtaining the local topography, we switched to MFM. The MFM signal was first acquired with a multi-domain, closed flux tip, where negligible (out-of-plane oriented) stray field should interact with the sample magnetic domains. Indeed, Figure \ref{fig:MFM}b shows that no MFM signal could be measured below the noise level of \SI{1.5}{\milli\hertz}. The lack of topography cross-talk in the lift mode image also demonstrates that the lateral variation of the electrostatic force is neglible. 

Following, the tip was pulsed by a \SI{160}{\milli\ampere} current pulse for \SI{205}{\nano\second}, which aligned the tip domain in the downward position, as indicated schematically in Figure \ref{fig:MFM}c and confirmed with Kerr microscopy prior. For safety, the tip was retracted by \SI{0.5}{\micro\meter} from the surface during the pulse, which can give rise to lateral drift of around \SI{10}{\nano\meter} in our SPM at these temperatures. With the magnetically oriented tip, we continue MFM imaging at \SI{100}{\kelvin} and observed a complex landscape of magnetic textures across the scanned area, as given in Figure \ref{fig:MFM}c. The smallest magnetic objects are highlighted with red circles in Figure \ref{fig:MFM}c, which also correspond to the strongest attractive magnetic forces. These features have on average a diameter of \SI{10}{\nano\meter}. We attribute these area's as stray field induced magnetised domains. Observing the nominal size, it is very likely that these corresponds to the weak FM textures \cite{Anahory2016EmergentInterfaces}, even at \SI{100}{\kelvin}. 

Performing force-distance (F-z) spectroscopy on the red islands of Figure \ref{fig:MFM}d showed complex behaviour and provides more evidence for weak FM properties. The tip was retracted  up to \SI{20}{\nano\meter} above the surface, and then lowered until a notable repulsive fore was observed. The frequency shift $df$ was measured during spectroscopy. Evidence of the tip stray field induced magnetic alignment is given in Figure \ref{fig:MFM}d.  We observe four distinct regimes; firstly we note a long-range attractive forces between \SI{20}{\nano\meter} and \SI{8}{\nano\meter}. This can be assigned to long range electrostatic forces. At around \SI{7.5}{\nano\meter}, a sudden negative change in frequency (force) is observed as indicated with a blue arrow. We attribute this to the significant increase of the magnetic field the sample experiences as the SM-PP tip approaches the weak FM domain and hence saturating it. At \SI{3.4}{\nano\meter}, indicated with an orange arrow, the attractive van der Waals force region is noted. At very small tip-sample distances the frequency shift becomes positive evidence of repulsive forces. We also measured the damping, a sign of energy loss via local magnetization change of the weak FM islands \cite{Torre2011MagneticNanoparticles}. In Figure \ref{fig:MFM}d, the red curve shows a sudden rise in the damping as indicated with the red arrow. Likely, at this distance the weak FM islands are magnetized periodically as the tip oscillates up and down. As a comparison, Figure \ref{fig:MFM}e shows the same F-z spectroscopy experiment performed on the blue areas of Figure \ref{fig:MFM}c. Less perturbing of the attractive force is noted, and no measurable dissipation change is observed as highlighted with the colored arrows. We conjecture that those blue colored areas are the antiferromagnetic domains. 

Generally, the weak FM features (coloured red) are embedded in magnetic labyrinth-like domains colored yellow/green in Figure \ref{fig:MFM}c. These domains are continuous and spread across the surface. Furthermore, they have smaller attractive force than the weak FM areas. Due to the tip's large stray field induced magnetization of LMO$_3$, no repulsive area's could be observed. Areas depicted in blue are observed in two distinct regimes. First, we note distributed areas highlighted with the dashed white line. Secondly, blue circular like objects are noted as highlighted with the blue circle. These objects show very little attractive frequency shift. Hence, excluding electrostatic forces as these are constant across the surface, these domains form an antiferromagnetic texture, corroborating the SSM observations of Wang \textit{et al.} \cite{Wang2015ImagingHeterostructures} and Anahory \textit{et al.} \cite{Anahory2016EmergentInterfaces}.

Furthermore, we note that the tip stray field can induce local magnetic perturbations in the real-space imaging. By comparing the forward and the backward scan, Figures \ref{fig:MFM}f-g, the areas highlighted in the black circle show clear distinction between the two images. We conjecture that the field from the tip perturbed the local weak FM domains. This would also be in agreement with the observation of streak-lines as indicated with the arrows. 

In conclusion, the results lay strong indications of the imaging capabilities of the magnetically controllable SM-PP tips for weak FM islands with a resolution higher than \SI{10}{\nano\meter}. Firstly, we achieved a repeatable control over the magnetization at the SM-PP tip with a consistently distributed domain state at the tip apex. Following, we demonstrated imaging of a complex magnetic texture of the rare-earth metal oxide perovskite LMO$_3$ with nanometric identification of weak FM islands. For further investigation of LMO$_3$ the SM-PP can be employed for ultra-high resolution imaging of the local u.c. variation in film thickness and the possible correlation with weak FM islands. Furthermore, the integration of the SM-PP in a LHe cryostat would increase the Q-factor by another order of magnitude, significantly improving the signal-to-noise ratio. Finally, future application the SM-PP can be combined with scanning tunneling microscopy functionality because of the tip metallic layers and electrode accessibility of the tuning fork. This way we can combine ultra-high lateral resolution imaging of conductive metal-oxide-perovskites and measure the long range MFM forces without the need to switch to different setups. This possibility opens up a approach to disentangle the atomic scale structure and long range magnetic ordering relations of transition metal oxides for application in spintronic and catalytic devices. Considering its enhanced sensitivity, the widened scope of tip-on-chip design can convert the MFM/AFM from a surface analysis tool with passive probes to a more sophisticated device with active more complex probes for characterization, e.g. nitrogen-vacancy centers diamond tips as quantum sensors for detecting ultra-small magnetic fields \cite{Casola2018ProbingDiamond, Healey2023QuantumHeterostructures} or currents \cite{Ariyaratne2018NanoscaleDiamond}.  

\subsection{Materials and Methods}

\subsubsection{Planar probe fabrication}
The metallic layers was sputtered on a thin \SI{150}{\micro\meter} thin Si $<100>$ wafer (intrinsic, UniversityWafer). The wafer was cleaved by a diamond scriber in \SI{1}{\square\milli\meter} pieces and inspected by an optical microscope. Following, the tip apex radius was inspected and selected for a radius below \SI{50}{\nano\meter} by a ZEISS-Sigma SEM. A FEI Nova600i SEM-FIB was used to fabricate the bridge structure by Ga-ion etching. A sequential beam current of \SI{0.05}{}, \SI{0.46}{} and \SI{2.8}{\nano\ampere} was used. Near the bridge the smallest current prevents damage and increases the etching resolution. The acceleration voltage was \SI{30}{\kilo\volt}.  The planar probe was placed onto the QTF (AB38T) prong with UV-curable resin with minimal volume (less than \SI{100}{\micro\liter}) employed by the use of a syringe needle. Silver paste was used to connect the electrical leads of the planar probe to those of the QTF. EPO-TEK 4410 was used to connect the wires from the QTF to a custom PEEK sensor holder. Detailed fabrication procedure is further outlined in Supplementary S1.

\subsubsection{Kerr Microscopy}

A Zeiss Axio Imager.D2m Kerr microscope was used with a $50\times$ magnification lens, assembled by Evico Magnetics with a polariser/analyser pair and manual slit diaphragm. The setup was combined with a set of water cooled Helmholtz coils for magnetic moment alignment by in-plane magnetic fields with respect to the Co film orientation. Kerrlab software was used for data acquisition. 
\subsubsection{Numerical calculations}
MuMax$^3$ was employed to simulate the domain structure of a \SI{16}{\nano\meter} Co film.  Exchange length of \SI{5}{\nano\meter} and grid unit cell of 4x4 \SI{}{\square\nano\meter} were used. For study of the thermal and magnetic properties of the SM-PP COMSOL Multiphysics was used with the AC/DC module and Heat Transfer module. Further numerical details are outlined in Supplementary S4. 

\subsubsection{Imaging in UHV}
A Scienta Omicron VT-SPM setup was modified to carry two additional electrical contacts for pulsing the planar probe tip. The contacts are constructed from 2 gold coated pogo pins placed on a custom PEEK holder onto the scanning tube. An square wave generator (Agilent 33120A) was connected to a custom MOSFET circuit to reduce the pulse down to several hundred ns. Coax cables where used to connect the function generator to the VT-SPM. Detailed imaging methods are outlined in Supplementary S5.


\begin{acknowledgement}

The authors thank W. Dijkstra for assistance in both the modification of the UHV-SPM and fabrication of the custom pulse generator. Special thanks to H. Hilgenkamp of Twente University of Technology, Netherlands, for providing the 6 u.c. LaMnO$_3$ thin film on SrTiO$_3$ sample. O. Kurnosikov acknowledge support from ANR-15-IDEX-04-LUE CAP-MAT ans by the “FEDER-FSE Lorraine et Massif Vosges 2014–2020. Financial support from the Eindhoven University of Technology is acknowledged. 

\end{acknowledgement}


\subsection{Supplementary S1: Fabrication and characterisation of the SM-PP}
\label{Supp:fabrication}

\begin{figure*}
    \centering
        \includegraphics[scale=0.45]{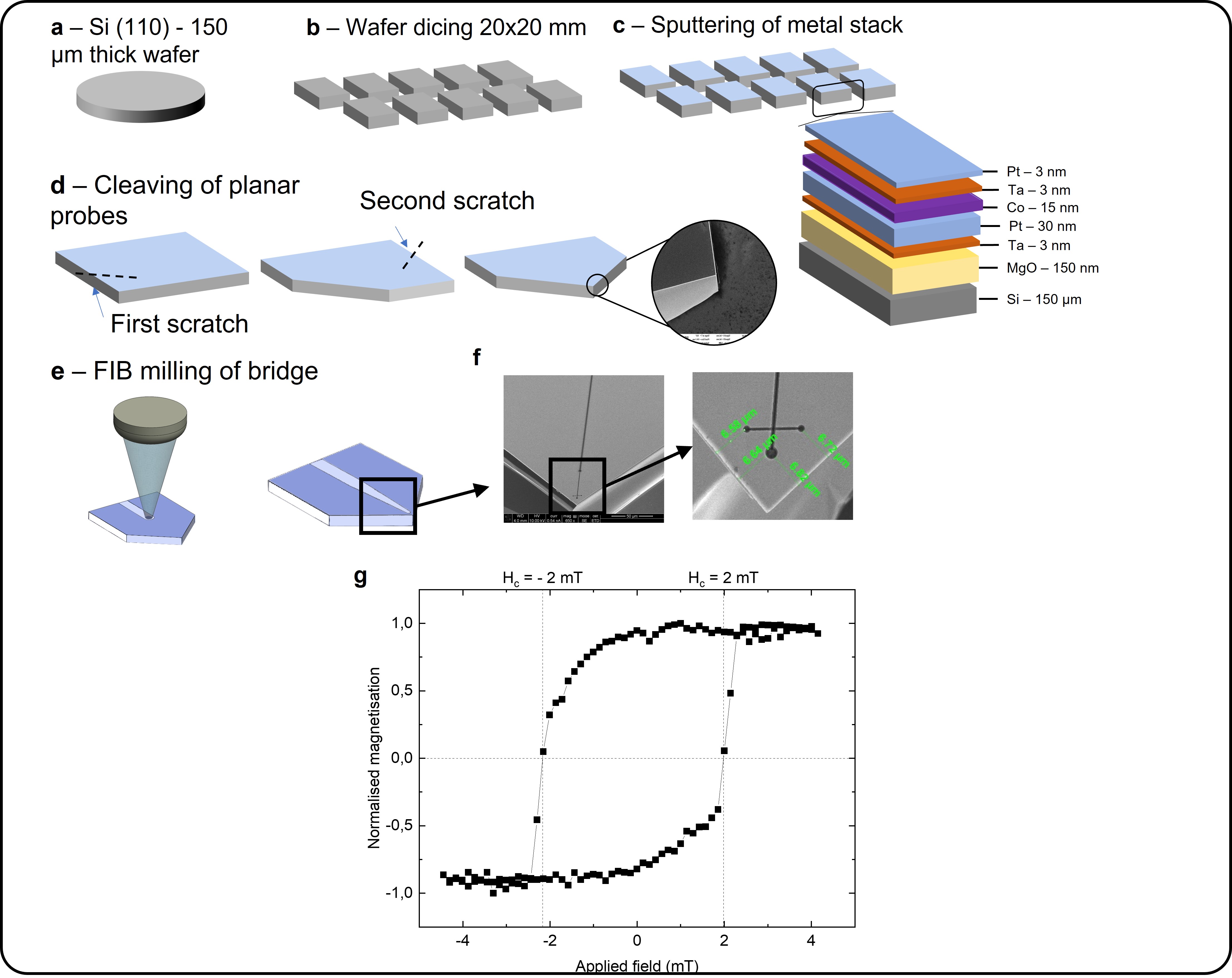}
            \caption{\textbf{Fabrication procedure of the switching-magnetisation planar probe.} (\textbf{a}) A \SI{150}{\micro\meter} thin intrinsic (110) silicon wafer is diced into smaller pieces, (\textbf{b}). (\textbf{c}) The pieces of wafer are sputtered with a \SI{150}{\nano\meter} MgO layer for electrical insulation needed to reduce the doping effect of Ga-ion implantation by FIB milling. Following, the metallic layers are sequentially deposited by plasma sputtering deposition, as schematically drawn. (\textbf{d}) After film deposition, the wafer pieces are cleaved into small rectangular planar probes with \SI{90}{\degree} angles forming the SPM tip. (\textbf{e}) Finally, the probe is functionalised with a bridge fabricated by FIB milling. (\textbf{f}) SEM images of a FIB fabricated bridge structure. (\textbf{g}) The magnetisation curve of the thin Co film on a planar probe as measured with Kerr microscopy.  }
                \label{fig:supp_fabrication}
\end{figure*}

The fabrication procedure of the SM-PP is schematically given in Figure \ref{fig:supp_fabrication}. The use of a thin \SI{150}{\micro\meter} Si wafer (intrinsic, from UniversityWafer) makes it possible to facilitate easy mechanical cleaving without having to apply large mechanical force and simultaneously reduce the planar probe's mass. First the wafer is diced into \SI{20}{}x\SI{20}{\square\milli\meter} pieces, as shown in Figure \ref{fig:supp_fabrication}b.  Single crystal silicon (100) is known to cleave in atomically smooth planes \cite{Lei2012DieReview}. By cleaving in two perpendicular directions a nanometer scale tip apex can be achieved. The cleaving results in square pieces up to \SI{1}{}x\SI{1}{\square\milli\meter}, with the tip apex' are evaluated for their radius (sharpness) with SEM, see Figure \ref{fig:supp_fabrication}e for a large scale image. For integration into a sensor, tips with a radius below \SI{50}{\nano\meter} were chosen for further fabrication steps, tips with larger radii where discarded. Because of the square nature of the cleaved planar probes, each piece offers up to \SI{4}{} adequate tip apices. This makes the availability of many excellent tips with a small radius very likely, fabricated in a short amount of time. 

With FIB milling, significant Ga-ion implantation occurred into the Si wafer which shortened the milled trench and voiding the bridge functionality. Hence, we resorted to growing an insulator spacer layer of SiO$_2$ or MgO between the metallic stack and the wafer. The integration of a microscale current pathway requires to reduce thermal-induced damage by excessive Joule heating. To enhance thermal management near the tip apex a high thermal conductivity MgO spacer layer is sputtered on top of the silicon wafer. MgO has a sufficient thermal conductivity of about \SI{40}{\watt\per\meter\per\kelvin} \cite{Slifka1998ThermalMeasurements}. Furthermore, MgO simultaneously facilitates electrical insulation to reduce electrical leakage currents. In Supplementary S4 we discuss the thermal dissipation behaviour between Si/MgO and Si/SiO$_2$ substrates after sending a current pulse through the bridge. 

\begin{figure*}
    \centering
        \includegraphics[scale=0.8]{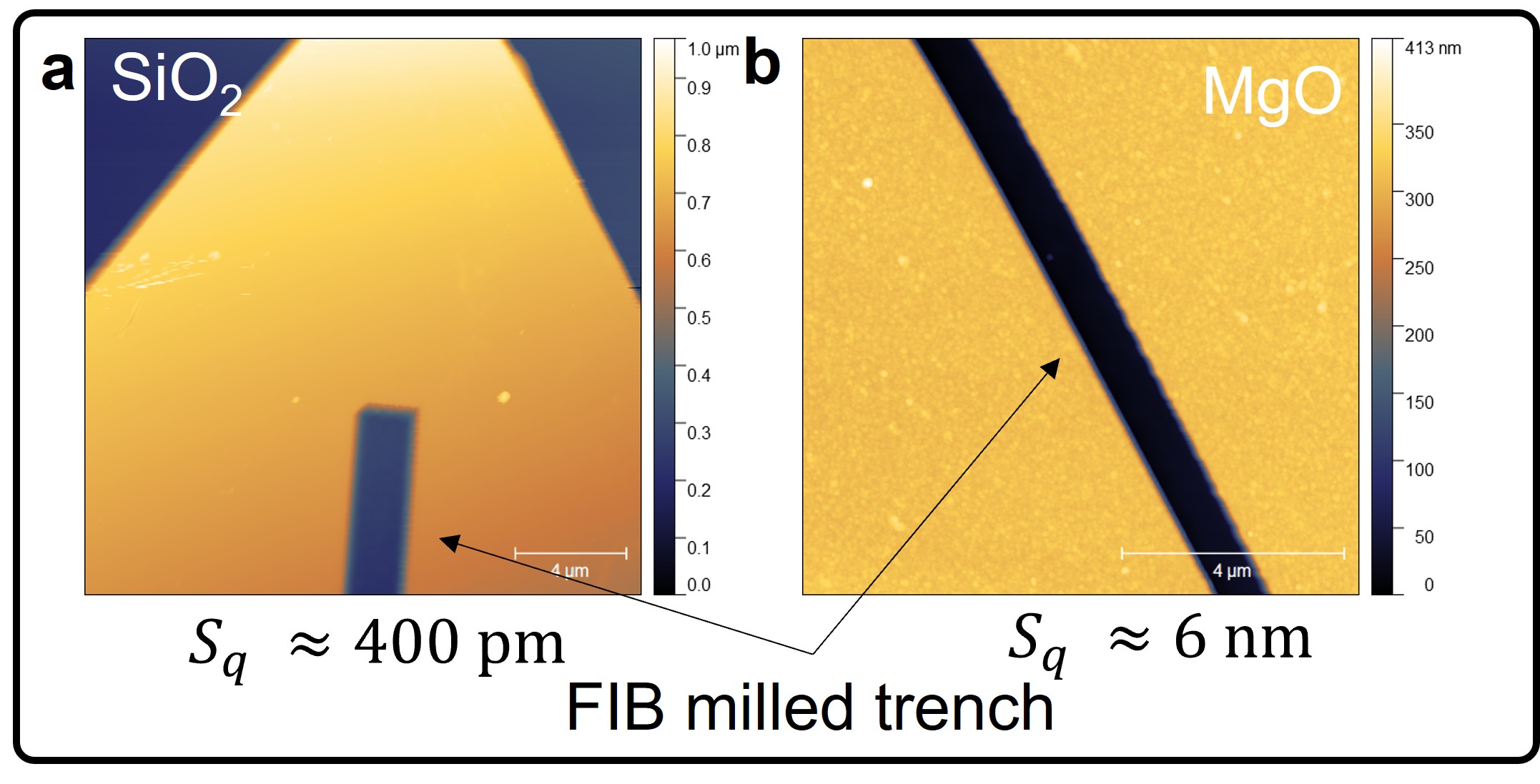}
            \caption{\textbf{Surface roughness of planar probe tips measured with AFM.} (\textbf{a}) A topographic AFM image of SiO$_2$ layered planar probe covered with the multi-layer metallic stack. The roughness is found to be around \SI{400}{\pico\meter}. (\textbf{b}) A planar probe surface but with a \SI{100}{\nano\meter} MgO layer instead of SiO$_2$. The surface roughness is larger compared to (\textbf{a}) and around \SI{6}{\nano\meter}. The black arrows point to the FIB milled trenches.}
                \label{fig:supp_roughness}
\end{figure*}

Next, the metallic multilayer was deposited. The planar probe metallic layer consists of the following structure, see Figure \ref{fig:supp_fabrication}c. First, a \SI{4}{\nano\meter} tantalum (Ta) seed layer is grown to induce good mechanical adhesion of the subsequent metal layers with the MgO/Si substrate. The Ta seed layer also smooths the surface roughness of the MgO layer to to some extent, which still results in a final RMS roughness of \SI{6}{\nano\meter}. Such a reduced roughness can actually be beneficial as it can be expected that near the cleaved tip apex small nanometer scale bumps form the nano-tip and reduce the van der Waals forces compared to a fully triangular structure, as it scales with the tip volume. The measured roughness of SiO$_2$ and MgO layered films are given in Figure \ref{fig:supp_roughness}. 

Subsequently, a \SI{30}{\nano\meter} current-carrying Pt layer is grown. This relatively thick Pt layer has the lowest film electrical resistance of the metallic stack; the majority of the current will flow through this layer. The ferromagnetic film is made from \SI{15}{\nano\meter} Co. Finally, the stacking is capped with \SI{3}{\nano\meter} of Ta and \SI{3}{\nano\meter} of Pt to induce high mechanical rigidity of the tip apex and prevent native oxidation.

\begin{figure*}
    \centering
        \includegraphics[scale=0.46]{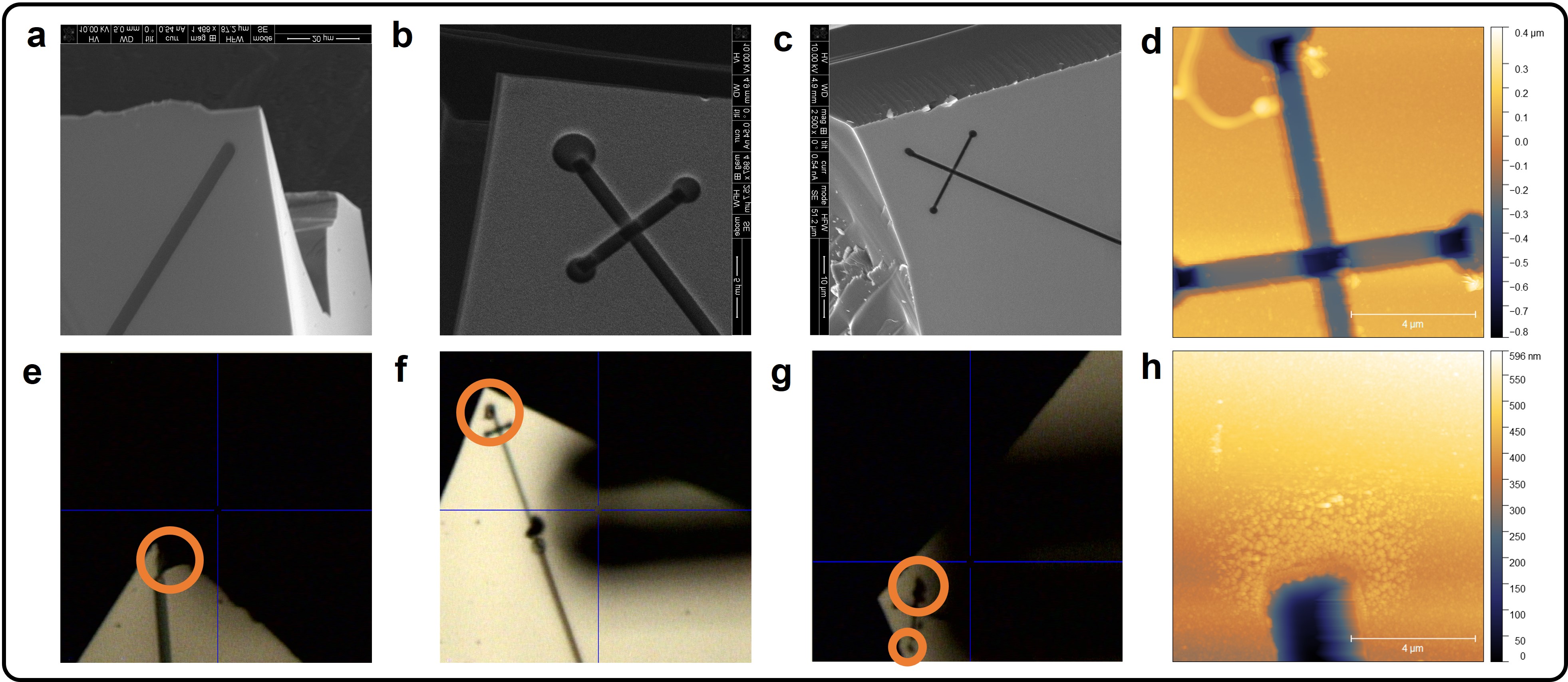}
            \caption{\textbf{SEM and AFM images of Joule heating-induced bridge damage with a SiO$_2$ layer.} (\textbf{a}) - (\textbf{c}) SEM image of pristine tips with FIB fabricated bridges. (\textbf{e}) - (\textbf{g}) Same tips after consecutive current pulsing of \SI{80}{\milli\ampere} showing film degradation. (\textbf{d}) and (\textbf{h}) AFM topography shows signs of film degradation and "peal-back" from excessive heating.}
                \label{fig:supp_damage}
\end{figure*}

When using (native) SiO$_2$ as a spacer layer between the intrinsic silicon substrate and the metallic stack of the planar probe, the low thermal conductivity of SiO$_2$ of only \SI{1}{\watt\per\meter\per\kelvin}, limits the thermal durability of the device. This low thermal conductivity was found to be insufficient to prevent damage from Joule heating to the bridge when using pulses above \SI{80}{\milli\ampere}. Limiting the current below \SI{80}{\milli\ampere} proved insufficient for full domain reversal for many devices with bridges in the micrometer widths. Figures \ref{fig:supp_damage}a, b and c show SEM images of FIB fabricated tips with either a single straight trench or the crossed configuration. The tips have a nominal bridge width, measured from the end of the trench to the tip end, of (a) \SI{12}{\micro\meter}, (b) \SI{10}{\micro\meter} and (c)  \SI{8}{\micro\meter}. Figures \ref{fig:supp_damage}e, f and g show optical microscope images of observed tip damage, highlighted with orange circles, after sequential \SI{80}{\milli\ampere} pulsing.  In these images, the metallic films have clearly degraded by excessive Joule heating. The Joule heating is most intense where the bridge width is smallest i.e. near the tip end, corresponding to the highest current density. Figures \ref{fig:supp_damage}d and \ref{fig:supp_damage}h show AFM topographic images of the FIB trench end, after 2 pulses. Clearly, the metallic film shows first signs of degradation or "peel-back", before full layer destruction occurs. With the inclusion of MgO, which is directly sputtered on top of the silicon wafer, we observed no Joule heating induced damage. MgO has a much high thermal conductivity of \SI{40}{\watt\per\meter\per\kelvin} \cite{Slifka1998ThermalMeasurements}. MgO devices pulsed over 25 times showed no degradation or change of the bridge resistance, even for pulses up to \SI{250}{\milli\ampere}.  

\subsection{Supplementary S2: Kerr microscopy}

A Zeiss Axio Imager.D2m Kerr microscope was used with a $50\times$ magnification lens, assembled by Evico Magnetics with a polariser/analyser pair and manual slit diaphragm. The setup was combined with a set of water cooled Helmholtz coils for magnetic moment alignment by in-plane magnetic fields with respect to the Co film orientation. Kerrlab software was used for data acquisition. The slit diaphragm makes it possible to filter light and select a magnetisation direction to visualise (horizontal or vertical) moment sensitivity. A custom-made sample holder was fabricated with integrated electrical wiring. The holder offered 3-degree's of positioning freedom needed for positioning of the SM-PP below the Kerr lens. A custom pulsing circuit was used to pulse a \SI{50}{\milli\ampere} to \SI{300}{\milli\ampere} current for \SI{150}{\nano\second} to \SI{500}{\nano\second}. By ramping up the current in consecutive pulses, the domain switching threshold was found. 

\begin{figure*}
    \centering
        \includegraphics[scale=0.42]{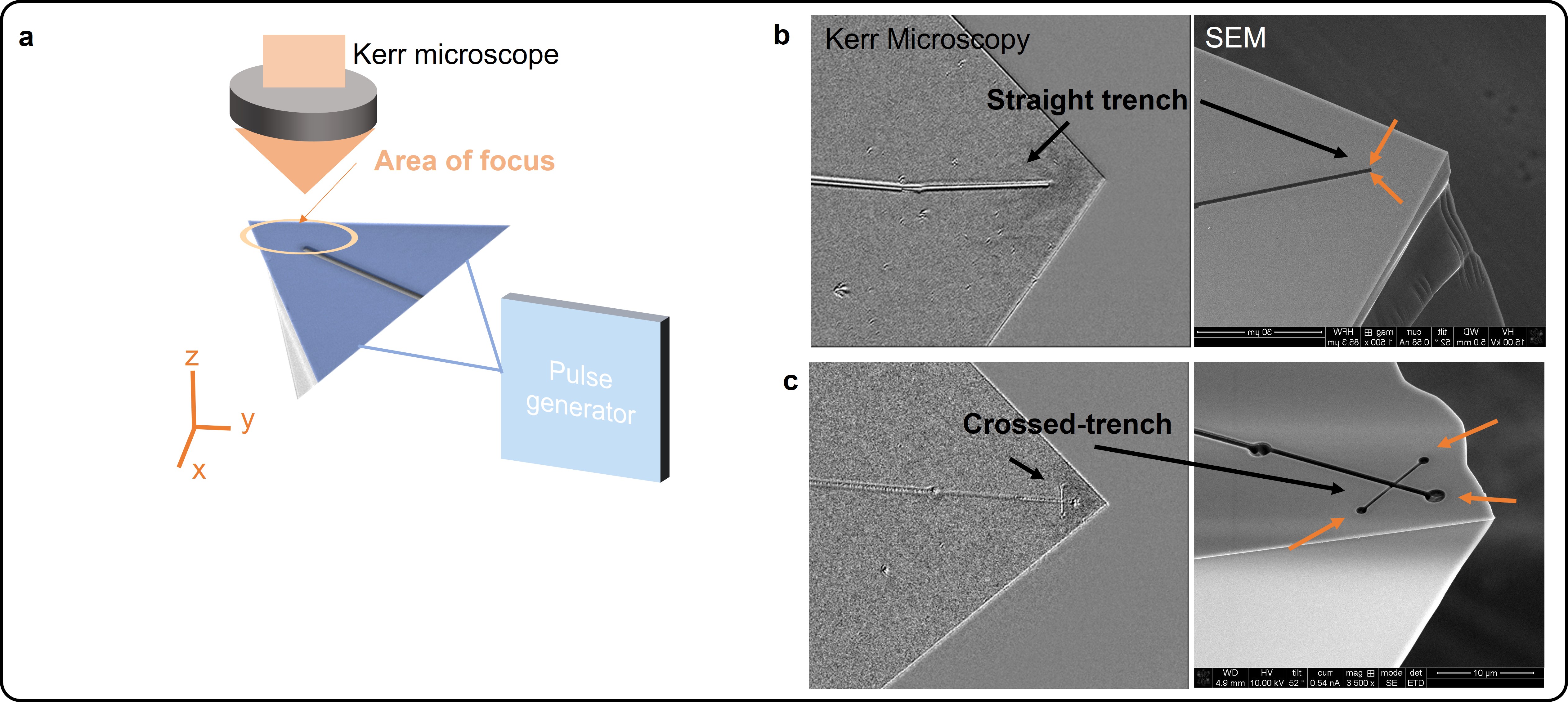}
            \caption{\textbf{Kerr microscopy.} (\textbf{a}) Setup to measure the tip magnetisation of a SM-PP device. The SM-PP is placed on a movable holder, to align the tip with the lens of the microscope. (\textbf{b}) and (\textbf{c}) show SEM images of FIB fabricated tips either with a straight trench, or with a crossed configuration. The black arrows highlight the pulsed tip Co domain, with a more stable domain with a crossed structure. The orange arrows  point to the bridge section of smallest diameters. These regions have the highest current density under pulsing. }
                \label{fig:supp_Kerr}
\end{figure*}

It was found that fabricating a single straight FIB trench does not always result in a stable magnetic domain reversal, with an example given in Supplementary Figure \ref{fig:supp_Kerr}b. For these straight trench devices, a stable domain was observed only after several consecutive current pulses. Even then, the domain does not extend across the complete bridge region, as indicated with the black arrows. With a single trench structure, only two nucleation sites are formed, near the very tip apex where the current density is highest (orange arrows). Although our numerical model suggests that it is possible to fully switch the domain with a single pulse, we attribute some devices needing more pulses to the fact that the FIB bridge was not always placed perfectly placed along the mirror-symmetrical axis of the probe. Hence, the current distribution is not equal along the bridge which could hamper full domain reversal near the tip apex. In our numerical models we have always assumed fully symmetrical structures.

To remedy this problem a perpendicular and smaller trench was FIB milled near the original trench, close to the tip apex, see Figure \ref{fig:supp_Kerr}c. This design adds an extra nucleation site, indicated with orange arrows, in close proximity to the bridge end. It was observed that this enables formation of a very stable domain that can be reliably reversed with a single pulse in all fabricated tips (number of tips exceeding 15). Figure \ref{fig:supp_Kerr}c show gray-scale Kerr Microscope images focused on the crossed bridge design. Kerr sensitivity was selected which corresponds to the domain pointing in-plane. Thus, the double crossed trench solution overcomes the non-perfect mirror-symmetric alignment of the FIB trench and induces multiple nucleation sites across the bridge.

\subsection{Supplementary S3: Note on the QTF function}

The large mass of the planar probe would decrease the Q-factor $Q$ below a few hundred, with $Q$ strongly influencing the sensitivity of the force sensor \cite{Giessibl2019TheMicroscope}. Hence, mass retuning of the QTF is performed by offsetting the mass of the planar probe carrying prong to equate that of the added mass of the oversized tip. This effectively restores $Q$ to its pristine value \cite{Ciftci2022EnhancingProbes}. The need to enhance $Q$ is vital by satisfying two requirements for the SM-PP to function. First, the frequency noise in frequency-modulation (FM) AFM scales inversely with $Q$
\cite{Giessibl2019TheMicroscope}. Hence, a higher $Q$ results in larger signal-to-noise ratio needed to measure the smaller forces associated with magnetic stray field gradient sensing \cite{Giessibl2019TheMicroscope}. 

Second, due to the mentioned separation of the electrode layout, the actual force-to-voltage conversion signal by the piezo-electric effect of the quartz tuning fork, is measured in the upper prong of the QTF. This prong essentially measures the tip-sample force indirectly and works most efficiently if the two oscillating prongs operate in the anti-phase mode with little dissipation in the connecting node \cite{Ciftci2022EnhancingProbes}. A low $Q$ would lead to insufficient coupling between the two oscillating prongs and hence impede operation of the force sensing. This is fundamentally different from the qPlus sensor, were only one prong oscillates and hence the force sensing is limited to this prong \cite{Giessibl2019TheMicroscope}.

Using the retuned tuning fork approach, we explore the possibility of using a high spring constant $k$ ($\sim \SI{e4}{\newton\per\meter}$) force sensors, which is remarkably higher than those of the qPlus (\SI{1500}{\newton\per\meter}). With a high $k$, the minimal detectable force gradient $dF_{\text{m}}/dz$ is reduced together with the detectable frequency shift down to several tens of mHz, according to the equation $\omega/\delta\omega = 1/(2k) \cdot dF/dz$ \cite{Giessibl2019TheMicroscope}. Hence, the tip needs to scan only a few nanometers above the sample surface to measure the magnetic stray field gradients, above the noise level of approximately \SI{1.5}{\milli\hertz}. This small $dF_{\text{m}}/dz$ simultaneously prevents the signals from areas beyond the dimensions of the tip to be picked up and hence the resolution can be increased significantly, needed for imaging  nanometer scale SP islands of LMO$_3$. Our SM-PP realises a $Q$ above \SI{20000}{} at room temperature, in ultra-high vacuum (UHV), which results in a noise floor of \SI{1.5}{\milli\hertz}.  For MFM imaging of nanometer sized SP textures, the high $k$ and large $Q$ are beneficial, as forces coming from area's not directly beneath the tip are not picked up by the SM-PP because the $df/dz$ decreases rapidly below the noise floor.

\subsection{Supplementary S4: Numerical calculations of thermal and magnetic properties}

\subsubsection{Influence of current pulse characteristics on Joule heating}

\begin{figure*}
    \centering
        \includegraphics[scale=0.42]{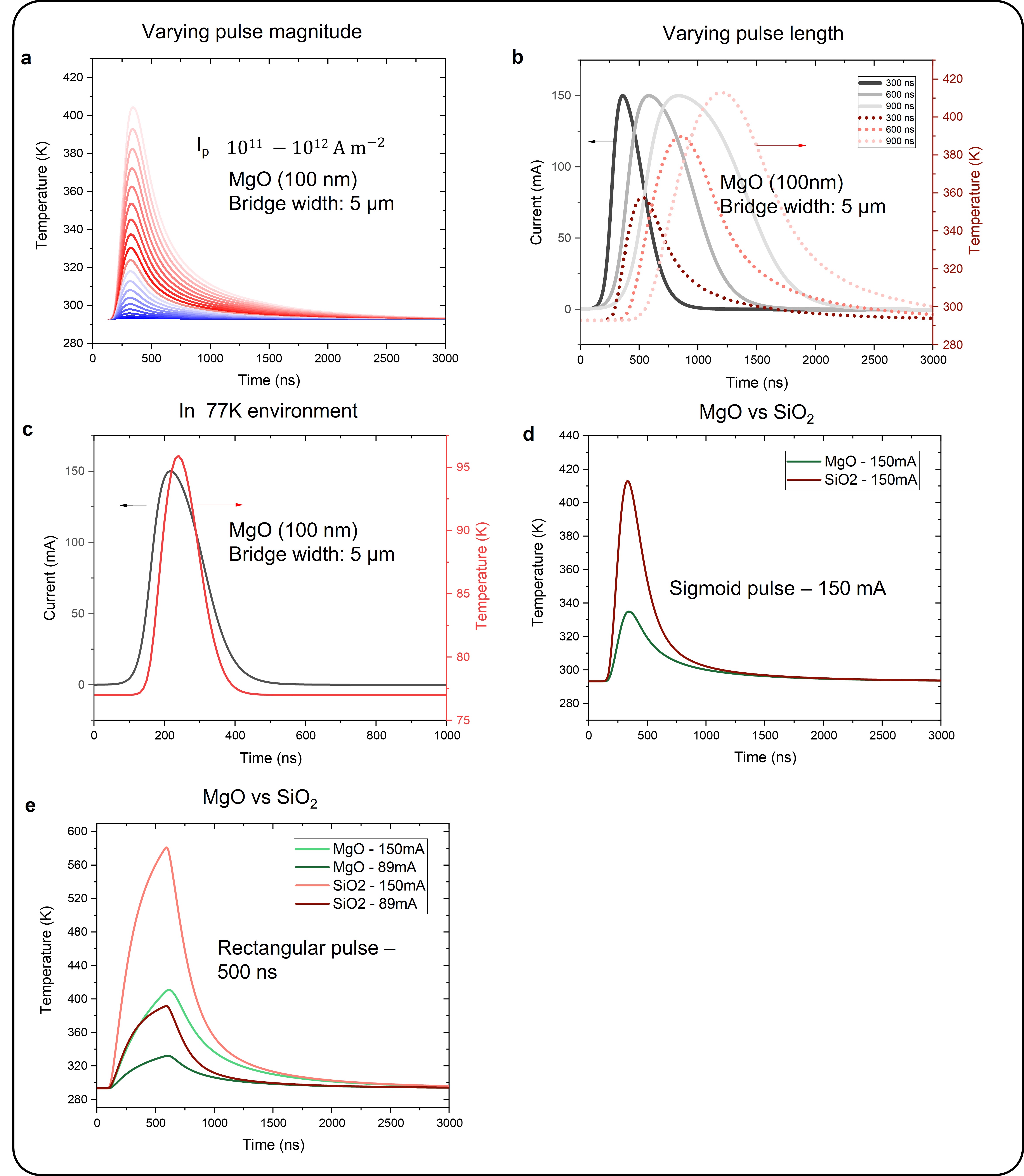}
            \caption{\textbf{Temperature response of the SM-PP for different pulse characteristics and spacer layer materials.} (\textbf{a}) Increase in temperature for different current densities. Even for very high currents exceeding \SI{e12}{\ampere\per\meter\squared}, the temperature does barely exceed \SI{400}{\kelvin}. (\textbf{b}) The temperature magnitude strongly depends on the pulse length between \SI{300}{} and \SI{900}{\nano\second}. (\textbf{c}) Operation in UHV at \SI{77}{\kelvin} should be possible since only a marginal increase in temperature is observed for a pulse sufficient to change the magnetisation of the SM-PP. (\textbf{d}, \textbf{e}) Simulation of the temperature response of MgO vs. SiO$_2$ spacer layers. The pulse shape is varied  between an asymmetric double sigmoid (\textbf{d}) and a square pulse (\textbf{e}), highlighting the need to keep the maximum current as short as possible to reduce thermal heating.}
                \label{fig:supp_Heating}
\end{figure*}

The (transient) temperature response of the SM-PP bridge was numerically studied with COMSOL \texttrademark. First we address the current pulse shape. 

In our SPM setup, capacitance of the cables that are connected to the SM-PP affect the pulse shape, resulting in a more asymmetric shape deviating from a square current pulse input. We simulate this curve in COMSOL using the asymmetric double sigmoidal function, which has the following functional form:

\begin{equation}
    I\left(t\right)=A_1/\left(1+\exp\left(-\left(t-t_c+w_1/2\right)/w_2\right)\right)\left[1-1/\left(1+\exp\left(-\left(t-t_c-w_1/2\right)/w_3\right)\right)\right]
    \label{eq:sigmoidal}
\end{equation}

Here, $w_1$ defines the width of the pulse, $w_2$ and $w_3$ together define the asymmetry of the pulse, $t_c$ is the centre of the pulse and $A$ is proportional to the amplitude of the current pulse $I_0$. We fit Eq 1 to the real pulse of which we obtain the full-width-at-half-maximum (FWHM), and pulse amplitude $I_0$.

Heat transport in ultrahigh vacuum (UHV) conditions is modelled via COMSOL\texttrademark   heat conduction. Thermal dissipation occurs via conduction within the planar probe and via radiative emission. The environment is set at room temperature, as is the case for the Scienta Omicron VT-SPM lacking a cryostat. Only the sample is cooled. The radiative thermal dissipation is set by the surface emissivity $\varepsilon$. We used the values for the following materials that were used in the functionalised planar probe: $\varepsilon_{\text{Si}}=0.6$, $\varepsilon_{\text{Pt}}=0.04$, $\varepsilon_{{\rm SiO}_2}=0.8$ and $\varepsilon_{\text{MgO}}=0.5$. The other metals, Ta and Co, are neglected. 

First, we consider the maximum current through the bridge  without overheating it. To this end, we varied the current density $J$ in the planar probe, which is related to the total current $I_p$ via
$J=\frac{I_{p}}{d_{Pt}d}$,
where $d_{Pt}$ is the thickness of the main platinum layer and $d$ is the width of the bridge gap (\SI{5}{\micro\meter}), respectively. The situation we consider is the asymmetric pulse from Equation \ref{eq:sigmoidal} with a current peak of \SI{150}{\milli\ampere}, which amounts to a current density of \SI{6e11}{\ampere\per\meter\squared}. We vary the current density from \SI{e11}{\ampere\per\meter\squared}  to \SI{e12}{\ampere\per\meter\squared}, in steps of \SI{0.5e11}{\ampere\per\meter\squared}. The results are shown in Figure \ref{fig:supp_Heating}a, in which the temperature barely increases by \SI{20}{\kelvin} for the smallest current densities, and increases by approximately \SI{115}{\kelvin} in case of the highest current density.

To characterize the temperature increase with respect to the current pulse length, we simulate three current pulses with a different pulse duration: \SI{300}{}, \SI{600}{} and \SI{900}{\nano\second} FWHM. The pulses are indicated in Figure \ref{fig:supp_Heating}b in dark solid lines. The resulting temperature evolution at the probe’s tip end is shown in Figure \ref{fig:supp_Heating}b in dotted red lines. Indeed, the temperature increases with the duration of the pulse. The temperature increases by about \SI{45}{\kelvin} in case of a short \SI{160}{\nano\second} pulse, up to an increase of over \SI{100}{\kelvin} for the \SI{900}{\nano\second} pulse. This large difference illustrates that the current pulse duration has a major impact on the temperature increase, so the current pulses need to be as short as possible in order to prevent overheating the probe.

We also investigated the thermal aspects of the planar probe when it is cooled down to liquid nitrogen temperature (\SI{77}{\kelvin}). The values of the surface emissivity are set to the same values as at \SI{293}{\kelvin}. The temperature evolution of the probe’s tip at \SI{77}{\kelvin} is shown in Figure \ref{fig:supp_Heating}c.  Evidently, the temperature of the probe increases by only \SI{35}{\kelvin}, which would make operation in a cryostat possible. 

Finally, we discuss the thermal characteristics of a SiO$_2$ layer. As Supplementary S1 discussed, the inclusion of SiO$_2$ as a spacer layer to reduce electrical shorting by FIB milling, turned out to negatively impact the thermal dissipation, eventually leading to bridge damage after several pulses above \SI{80}{\milli\ampere}. Numerical calculations of MgO and SiO$_2$ are shown in Figure \ref{fig:supp_Heating}d. These plots  show a temperature increase when SiO$_2$ is used, to a maximum above \SI{400}{\kelvin}. These values were calculated with a pulse of equation \ref{eq:sigmoidal}, with a FWHM of \SI{160}{\nano\second}. In the case of a \SI{230}{\nano\meter} MgO layer, the temperature increase is less than \SI{40}{\kelvin}, supporting the experimentally observed stability of MgO as a spacer layer. For comparison, we have also used a block pulse of \SI{160}{\nano\second} with a magnitude of \SI{150}{\milli\ampere} with the thermal response shown in Figure \ref{fig:supp_Heating}e, with much higher temperature peaks observed, exceeding over \SI{500}{\kelvin}. Hence, the peak value needs to be minimized to only a few ns to prevent damage to the tip, and a block pulse is not sufficient. 

\subsubsection{MuMax3 calculation of tip domain orientation}
MuMax$^3$ \cite{Vansteenkiste2014TheMuMax3} was used in which a \SI{16}{\nano\meter} Co film was simulated, with an exchange length of \SI{5}{\nano\meter} and grid unit cell of 4x4 \SI{}{\square\nano\meter}. The following values are taken to simulate the cobalt; saturation magnetisation $M_{sat}$ = \SI{1.4}{\mega\ampere\per\meter}, exchange constant $A_{ex}$ = \SI{16}{\pico\joule\per\meter} and the $1^{st}$ order uniaxial anisotropy constant $Ku_1$ of \SI{0.72}{\mega\joule\per\cubic\meter}.

\subsubsection{Note on the magnetic stray field computation}

\begin{figure}
    \centering
    \includegraphics[width=\columnwidth]{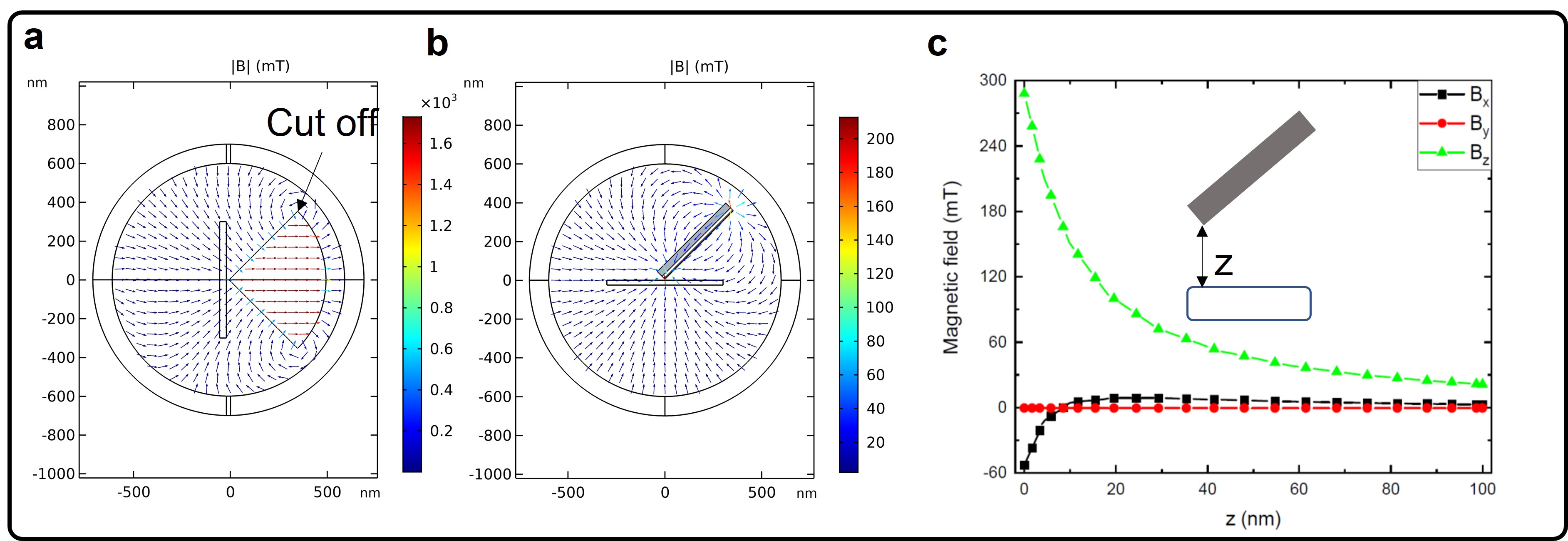}
    \caption{\textbf{COMSOL-calculated magnetic field distribution around the tip}. (\textbf{a}), (\textbf{b}) Arrow plot of the magnetic field distribution. Color bar gives the absolute strength in \si{\milli\tesla}. (\textbf{c}) Magnetic field components $B_x$, $B_y$ and $B_z$ as a function of tip-sample distance $z$. }
    \label{fig:Supp_MagneticFieldDistributions}
\end{figure}

To study the magnetic properties of the SM-PP, specifically its tip magnetic field component distribution $B_x$, $B_y$ and $B_z$ and the field magnitude, a 3D COMSOL\texttrademark  FEM model was made. The SM-PP is modelled as a \SI{15}{\nano\meter} ferromagnetic Co layer with a magnetization of \(M=\SI{1400}{\kilo\ampere\per\meter}\). In the main text we showed the stray field of the SM-PP in the in-plane direction of the tip as it extends outwards, when placed a certain height $z$ above a flat sample. In the calculations, the SM-PP and sample are placed in an air medium. Figure \ref{fig:Supp_MagneticFieldDistributions} presents the modeled stray field results.

Figures \ref{fig:Supp_MagneticFieldDistributions}a and b show the field distribution at two different cross-sections: in the $xy$-plane, and parallel to the probe surface, respectively. Important to note is that the probe length is much larger than the tip-sample distance $z$. However, for computational reasons the probe was made smaller ("cut-off"), which gives rise to an additional curvature of the field near the upper side of the probe. Near the tip, where the calculations were performed, the effect of this curvature has been taken into account during the final calculations. Figure \ref{fig:Supp_MagneticFieldDistributions}c presents the calculated $B_x$, $B_y$ and $B_z$ between the tip sample distance $z$. 

\begin{figure}
    \centering
    \includegraphics[width=\columnwidth]{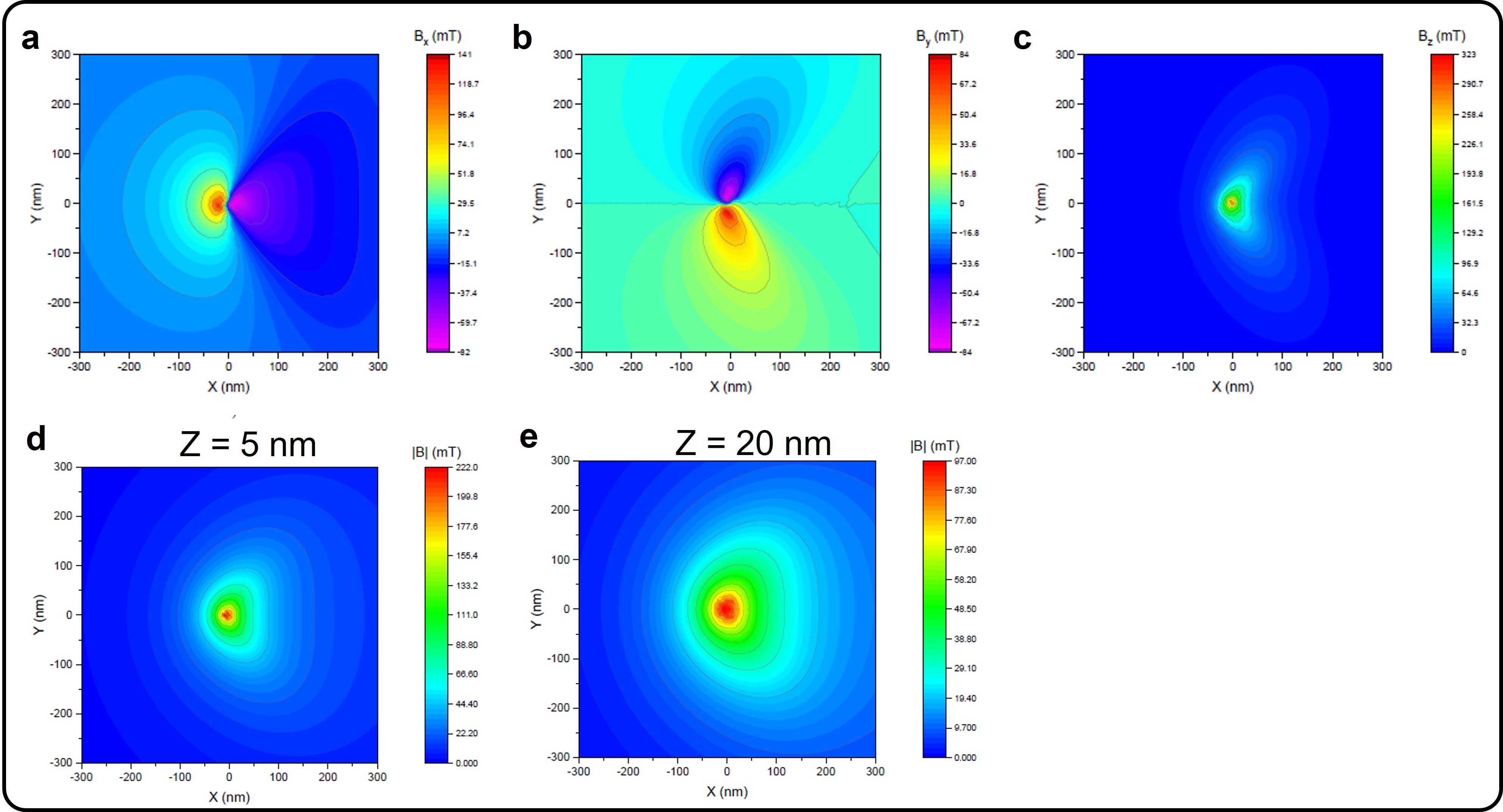}
    \caption{\textbf{Distribution of the SM-PP tip stray field on a sample surface in the xy plane.} (\textbf{a})-(\textbf{c}) Numerically calculated stray field components across the sample surface placed at z = 0 nm and the tip apex positioned at \SI{1}{\nano\meter}. (\textbf{d}),(\textbf{e}) Numerical calculated stray field  and spatial distribution at different heights of the tip (z)}
    \label{fig:Supp_MFM_Distribution}
\end{figure}

We calculated the spatial distribution of the tip's stray field $\vec{B}$ across the sample surface. Figure \ref{fig:Supp_MFM_Distribution} shows the lateral distribution of $B_x$, $B_y$ and  $B_z$  at a tip-sample distance of \SI{1}{\nano\meter}. We expect this to be the closest tip-sample distance during oscillation. At the lowest point of the amplitude extension with respect to the sample surface, the stray field perturbation should be strongest. The $B_x$ and $B_y$ components in Figure \ref{fig:Supp_MFM_Distribution}a and \ref{fig:Supp_MFM_Distribution}b show sign inversion close to the (0,0) point, where the tip apex is positioned. This is to be expected from the symmetry of the planar probe. The magnitude of $\vec{B}$ can become quite large up to \SI{100}{\milli\tesla}. However, at (0,0) the field is minimal for $B_y$. But for $B_x$ a sizable field up to \SI{80}{\milli\tesla} is observed directly below the tip. For a fabricated planar probe, the tip is never fully symmetric, nor is the bridge shape and it's position on the tip apex. The in-plane values are hence expected to be larger below the tip during experiments. The out-of-plane component $B_z$ is strongest at (0,0) and given in Figure \ref{fig:Supp_MFM_Distribution}c. The magnitude can reach up to over \SI{300}{\milli\tesla}. For measuring LMO$_3$, having both an in-plane and out-of-plane field is beneficial, as the SM-PP in-plane field magnetises the SP islands, while $B_z$ is used to read out the stray signal. This is similar to the SSM experiments of Anahory \textit{et al.} \cite{Anahory2016EmergentInterfaces} and supports our work on observing the complex SP islands.  

We also examined the dependency of the magnetic field magnitude $|B|$ and distribution on $z$, between \SI{5}{\nano\meter} and \SI{20}{\nano\meter}. The spatial distribution of $|B|$ is given in Figure \ref{fig:Supp_MFM_Distribution}d and \ref{fig:Supp_MFM_Distribution}e, respectively. Evidently, $|B|$ increases by almost a factor of $3$ as $z$ decreases down to \SI{5}{\nano\meter}. Hence, this may explain why we observe a sudden change (kink) in F-z spectroscopy, in Figure \ref{fig:MFM} in the main text, as the SP islands are magnetized and a sudden change in the magnitude of the attractive force occurs. 

\begin{figure}
    \centering
    \includegraphics[width=\columnwidth]{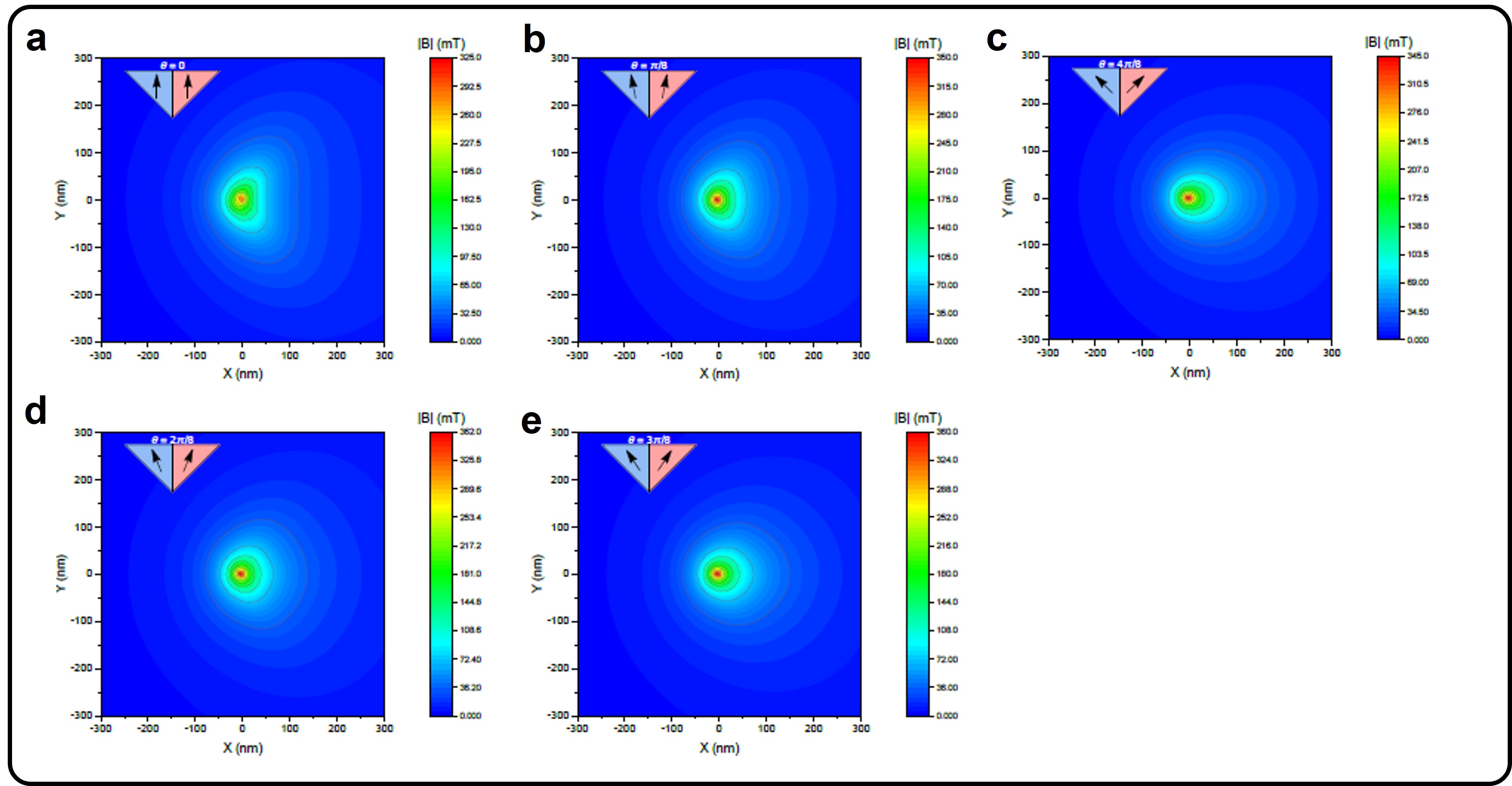}
    \caption{\textbf{Distribution of the SM-PP tip stray field on a sample surface in the xy plane, depending on the off-axis orientation of the in-plane magnetisation of the Co layer.}}
    \label{fig:Supp_Angle_Distribution}
\end{figure}

Finally, we examined the effect of the direction of the Co magnetisation along the planar probe symmetry axis, on the tip stray field magnitude and distribution. Figure \ref{fig:Supp_Angle_Distribution} shows the calculated $|B|$ at a fixed $z =\SI{0.5}{\nano\meter}$. Here, the relative Co magnetisation direction on both sides of the symmetry axis are depicted in blue and red with the angle between the two directions indicated as $\theta$. Two main conclusions can be derived from these results: the magnitude of magnetic field strength is relatively independent on $\theta$. And secondly, the shape of the magnetic field distribution on the surface changes only slightly. In conclusion, changing the symmetry of Co magnetisation along the planar probe vertical symmetry axis does not alter the final $|B|$ distribution a lot. Hence it is reasonable to assume $|B|$ is mainly tip-sample distance $z$, dependent.

\subsection{Supplementary S5: Methods SM-PP MFM experiment}
\subsubsection{LMO$_3$ sample in UHV}

The 6 u.c. LMO$_3$/STO$_3$ sample was exposed to ambient air prior to measurement, hence we expect a thin layer of contaminates present of the LMO$_3$ surface before being loaded into UHV. Gently heating to \SI{100}{\degree} in UHV removed the water adsorbents. Higher temperatures were not used to prevent oxygen diffusion which likely alters the stochiometry and magnetic behaviour \cite{Li2019ControllingEvidence, Xie2015EffectsFilms}. However, we cannot attest if small stochiometric changes have occurred (and also over time) or between sample growths. This variable offers future investigation possibilities between the stochiometry relationship and the magnetic textures. 

\subsubsection{FM-MFM imaging and data visualisation}

To measure the long-range MFM signal, the tip was retracted by \SI{5}{\nano\meter} and followed the previously obtained topography, the so-called lift mode \cite{Kazakova2019FrontiersMicroscopy}. For measuring the long range magnetic gradient force $dF_m/dz$ , the oscillation amplitude $A$ was set to \SI{10}{\nano\meter}.

A Scienta Omicron VT-SPM was used with a custom tip holder and low temperature sample holder for enhanced thermal conduction. Imaging was performed by locking the amplitude and tracking the frequency shift with a phase lock loop (PLL). Care was taken to keep all the scanning parameters constant, with optimized feedback settings for the phase, amplitude and frequency. The PLL bandwidth was set to \SI{384}{\hertz}. After a topography line-scan was obtained, the tip was lifted by the predefined lift height. Both forward and backward scans where compared to identify repeatability and exclude artifacts. We imaged with a speed of \SI{39}{\nano\meter\per\second} at a resolution of 256x256 pixels. Gwyddion software was used for the AFM and MFM data plotting and analysis. We used plane projection to level the image, and line-alignment in the vertical direction. Due to thermal drift induced by the thermal gradient between the cold sample and the room temperature SM-PP (as dictated by the VT-SPM design), a small eigenmode frequency shift of \SI{50}{\milli\hertz} was observed during the prolonged imaging, hence we renormalized the frequency MFM shift accordingly to the first few line-scans. No influence on the MFM lateral dimensions was noted by this effect. Finally, both the sample and tip are grounded. Future work will translate the SM-PP to a cryostat to reduce thermal gradients which would also significantly increase the $Q$ further. 

\subsection{Supplementary S6: Variable temperature MFM}

\begin{figure}
    \centering
    \includegraphics[scale=0.42]
    {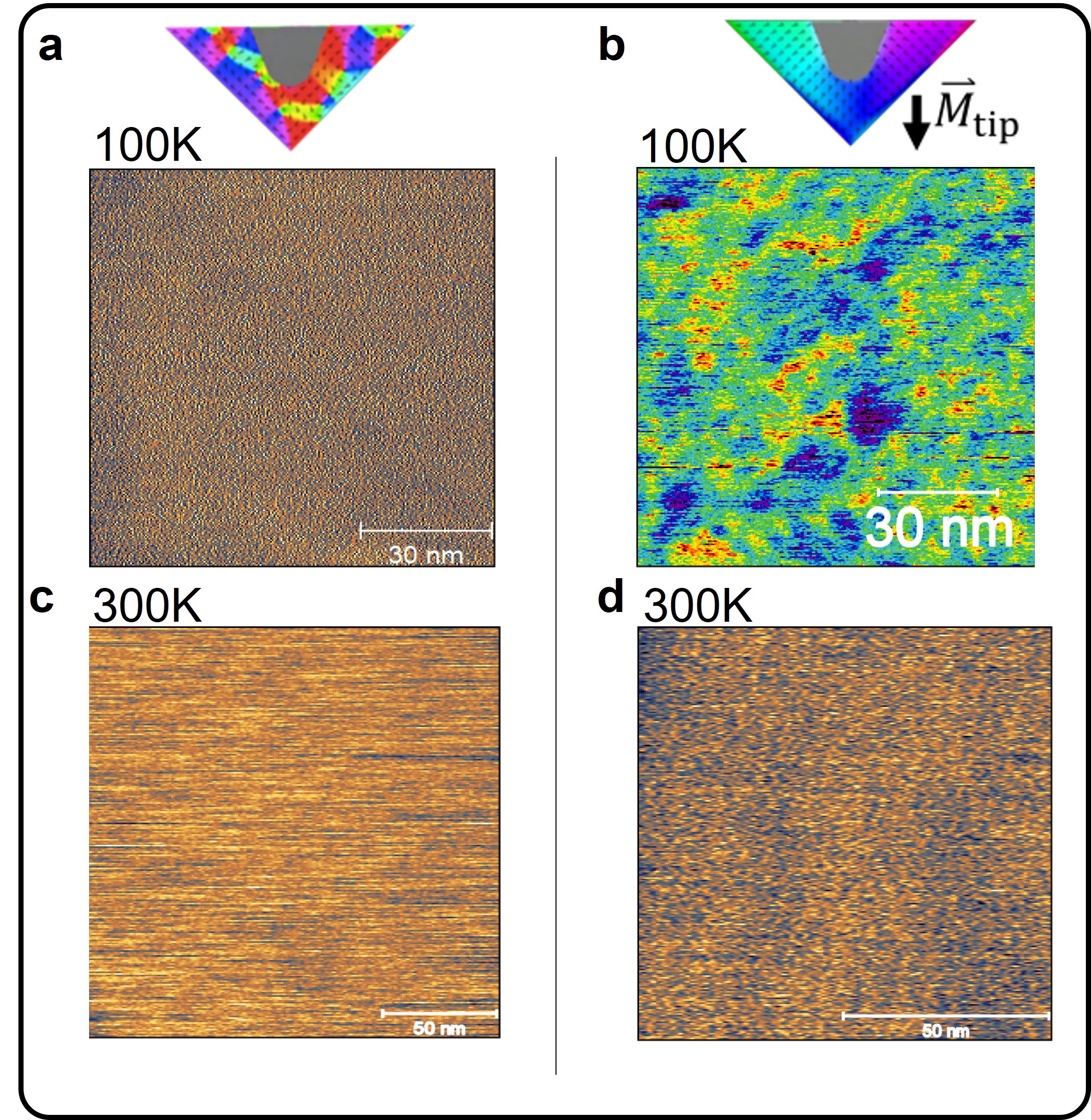}
    \caption{\textbf{MFM images at variable temperature. }(\textbf{a}), (\textbf{b}). MFM images taken with a multi-state and single-state domain at \SI{100}{\kelvin}, respectively. (\textbf{c}), (\textbf{d}) Multi-domain and single domaing MFM imaging at \SI{300}{\kelvin}, respectively.}
    \label{fig:Supp_VT_MFM}
\end{figure}


\bibliography{references.bib}


\end{document}